\documentclass[twocolumn,aps,prl,superscriptaddress,groupedaddress, nofootinbib]{revtex4}
\usepackage{textcomp}
\usepackage{amsmath}
\usepackage{amssymb}
\usepackage{graphicx, scalerel}
\usepackage{color}
\usepackage{mathtools}
\usepackage{natbib}
\setcitestyle{square, comma, numbers,sort&compress, super}

\usepackage{afterpage}
\usepackage{tikz}
\usepackage{tikzsymbols}
\usetikzlibrary{arrows.meta}

\usepackage{listings}
\usepackage{color}

\definecolor{dkgreen}{rgb}{0,0.6,0}
\definecolor{gray}{rgb}{0.5,0.5,0.5}
\definecolor{mauve}{rgb}{0.58,0,0.82}

\def\ear{\raisebox{-0.4ex}{\includegraphics[height=2.15ex]{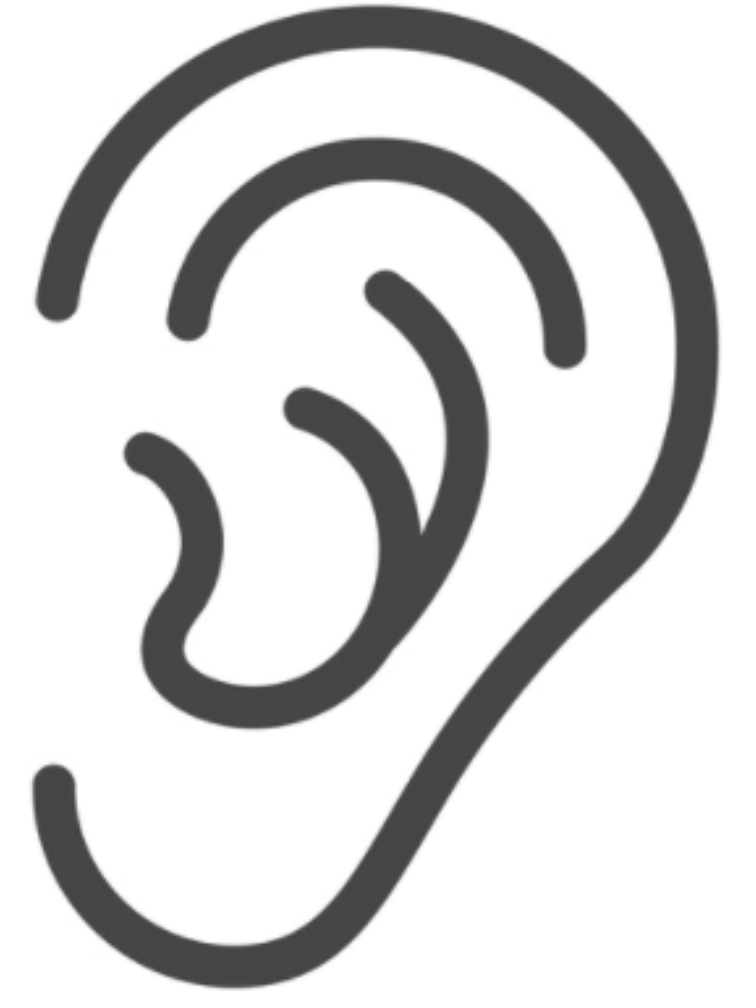}}
\newcommand\lowparen{\raisebox{\dimexpr-.5\ht\strutbox+.5\dp\strutbox}{)}}}

\def\catAlive{\raisebox{-0.4ex}{\includegraphics[height=2.15ex]{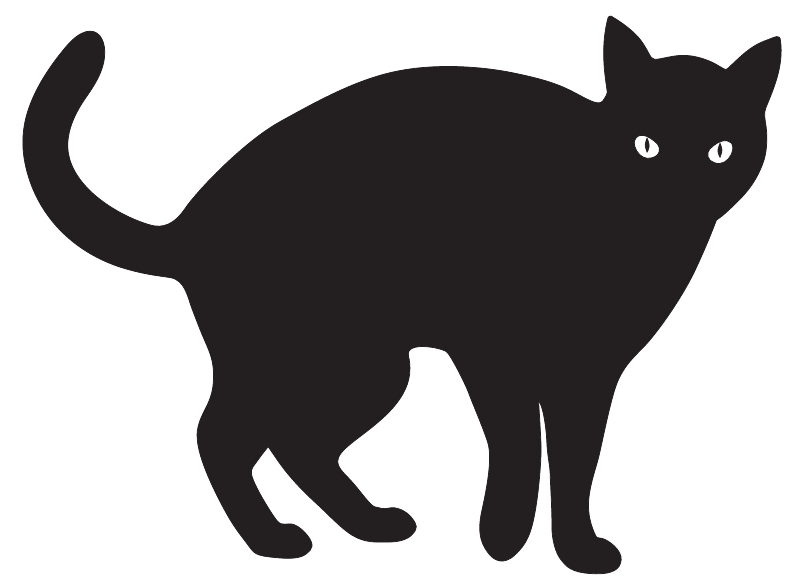}}}
\def\catDead{\raisebox{-0.4ex}{\includegraphics[height=2.15ex]{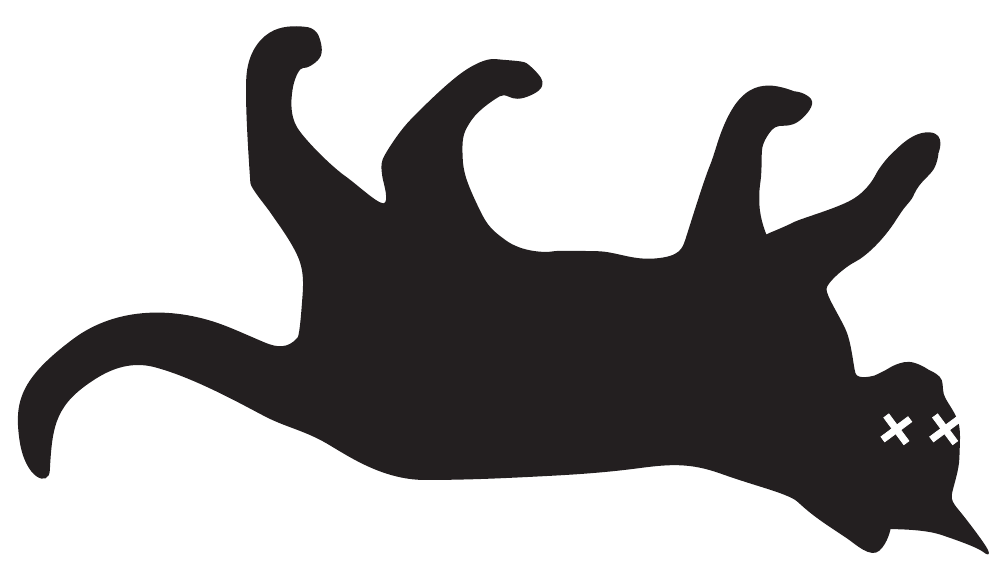}}}

\def\detectorRed{\raisebox{-0.3ex}{\includegraphics[height=2.15ex]{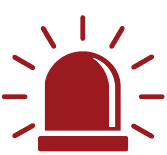}}}
\def\detectorBlue{\raisebox{-0.3ex}{\includegraphics[height=2.15ex]{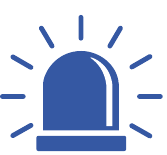}}}

\def \guy{
         \text{\raisebox{-0.4ex}{\Strichmaxerl[1.5]}}
         \hspace{-0.1cm}
         }

\definecolor{auburn}{rgb}{0.43, 0.21, 0.1}

\usepackage[toc,page]{appendix}

\usepackage{multirow}

\usepackage{blindtext}

    \newcommand*\lateraleye{%
       \raisebox{-0.4ex}{\scalebox{0.15}{
    \tikzset{every picture/.style={line width=0.75pt}} 
    \begin{tikzpicture}[x=0.75pt,y=0.75pt,yscale=-1,xscale=-1]
    \draw  [line width=1.5]  (300,100.33) .. controls (326,122) and (352,135) .. (378,139.33) .. controls (352,143.67) and (326,156.67) .. (300,178.33) ;
    \draw  [fill={rgb, 255:red, 0; green, 0; blue, 0 }  ,fill opacity=1 ] (308.94,116.33) .. controls (313.87,116.33) and (317.86,125.51) .. (317.85,136.83) .. controls (317.84,148.15) and (313.84,157.33) .. (308.91,157.33) .. controls (303.99,157.32) and (300,148.14) .. (300.01,136.82) .. controls (300.02,125.5) and (304.02,116.32) .. (308.94,116.33) -- cycle ;
    \draw  [draw opacity=0][line width=1.5]  (314.84,166.6) .. controls (311.87,164.64) and (309.14,162.18) .. (306.76,159.24) .. controls (295.12,144.82) and (296.6,124.33) .. (310.07,113.45) .. controls (311.48,112.32) and (312.96,111.33) .. (314.5,110.49) -- (331.14,139.55) -- cycle ; \draw  [line width=1.5]  (314.84,166.6) .. controls (311.87,164.64) and (309.14,162.18) .. (306.76,159.24) .. controls (295.12,144.82) and (296.6,124.33) .. (310.07,113.45) .. controls (311.48,112.32) and (312.96,111.33) .. (314.5,110.49) ;
    \draw  [fill={rgb, 255:red, 255; green, 255; blue, 255 }  ,fill opacity=1 ] (304.43,124.2) .. controls (306.09,124.25) and (307.32,128.01) .. (307.18,132.6) .. controls (307.05,137.19) and (305.59,140.88) .. (303.93,140.83) .. controls (302.27,140.78) and (301.03,137.02) .. (301.17,132.43) .. controls (301.31,127.83) and (302.76,124.15) .. (304.43,124.2) -- cycle ;
    \end{tikzpicture}
    }}\,}

\usetikzlibrary{positioning,calc}

\usepackage[nodisplayskipstretch]{setspace}

\usepackage{ntheorem}
\theoremstyle{empty}
\theoremheaderfont{\bfseries\upshape} 
\theorembodyfont{\itshape}
\newtheorem{named}{}

\begin{document}

\title{Falsifiable Tests for Theories that Govern How an Individual's Conscious Experience Traverses Everett's ``Many-Worlds'' Multiverse}

\author{Steven Sagona-Stophel}
\affiliation{Quantum Optics and Laser Science, 
Imperial College London, Prince Consort Rd, London SW7 2AZ, United Kingdom}
\affiliation{Corresponding author: stevensagona@gmail.com}

\begin{abstract}
We propose a set of simple quantum optics experiments that test for an entirely new domain of physical laws that govern how an individual's conscious experience traverses the multiverse within Everett's many worlds interpretation of quantum mechanics. These experiments imply an exception to the Born rule in a proposed ``observer-specific" reference frame. These experiments must be done by you, the reader. If it is performed by anyone else, other than you, the reader --  you will observe that the person performing the experiment will observe an outcome that is not special, interesting, or different from what is already known about quantum mechanics. To the best of our knowledge, this would be the first ever modern experiment that is only meaningful if the experiment is performed by you, the reader --  and cannot be inferred from the results of another experimenter. Therefore, since each individual must perform this test on his or her own, we outline a set of real experiments that can be easily performed such that as many people as possible can individually verify this for themselves. We do not know or specify what specific physical laws exist within this ``observer-specific'' domain, but come up with a number of different tests to cover as many theories as possible. 
\end{abstract}


\maketitle


Could there exist scientific knowledge that fundamentally cannot be discovered by experiments performed by others and can only be discovered by you, the reader?

There is at least one peculiar thought experiment that has this property, involving what is called “quantum immortality \cite{QuantumImmortality1}.” The idea of quantum immortality stems from the many worlds interpretation of quantum mechanics, a review of which is provided in SM I. 

“Quantum immortality” is the idea that observers, in their own perspective, cannot observe themselves dying. 

For example, if a real cat is prepared in a Schr\"{o}dinger’s cat state $\frac{1}{\sqrt{2}}(|\catAlive\rangle +  |\catDead\rangle)$, in which the cat is placed in a superposition of being alive $|\catAlive\rangle$ and dead $|\catDead\rangle$, the existence of quantum immortality implies that the cat, in its perspective, always survives and never experiences itself dying.

But to anyone opening the box, the cat is just as likely to be found dead as alive. Therefore, if quantum immortality is correct, the two following things are true after the box is opened: In the cat's perspective, there is a 100\% chance that it survives and a 0\% chance that it dies. In the external observer's perspective, there is a 50\% chance to be observed to be dead and a 50\% chance to be observed to be alive. 


Many of the subsequent discussions of quantum immortality \cite{ QuantumImmortality2, QuantumImmortality3, QuantumImmortality4, QuantumImmortality5} try to argue whether or not this is true or if it is sensible or rational to believe in quantum immortality, derived from first principles within the standard theory of quantum mechanics. 

Unlike these previous works, here we begin by immediately assuming that quantum immortality is \textit{not} consistent with the standard theory of quantum mechanics. Instead we propose that quantum immortality can be ``patched'' into quantum mechanics by adding an extra rule into quantum mechanics. 
 
We call this extra rule the ``no death rule'' which can be written as: ``an observer cannot experience itself in worldlines where that observer is dead.'' Discussed in more detail in SM II, we introduce a new notation to express this extra rule:
\begin{align*}
&\textbf{The No Death Rule:}\\
 &\begin{cases}
 &\scalebox{1.5}{Pr}\Big(\catAlive : |\catAlive\rangle \rightarrow |\catDead\rangle \Big) = 0 \\
 &\scalebox{1.5}{Pr}\Big(\catAlive : |\catAlive\rangle \rightarrow |\catAlive\rangle \Big) = 1 \\
 & \textbf{otherwise, Born Rule}\\
  \end{cases}
\end{align*}
\noindent 
The symbol probability $\scalebox{1.5}{Pr}$ represents the probability. The state to the left of the colon represents the perspective the observer making the measurement, while to the right of the colon represents which outcome that is experienced.

In this paper we call the ``no death rule''  an ``observer-dependent rule,'' which we define as a rule in which the probabilities of the observed outcome depends on the perspective of the observer. 

Like any scientific theory, modifying the theory of quantum mechanics usually produces a different theoretical outcome that can be tested in an experiment. However, the addition of the ``no death rule'' to quantum mechanics is not testable in a conventional experiment. This is because the observer-dependent ``no death rule'' specifies that in a Schr\"{o}dinger's cat experiment something different from normal quantum mechanics occurs \textit{only in the perspective of the cat}. Regardless of if this special ``no death rule'' exists or not, anyone performing an experiment \textit{by opening the box} will not observe any changes in the results of the experiment. The only way to test for this rule is by \textit{being the cat inside the box}.

\section{Main Idea}

Is it possible that a similarly compelling idea exists that does not require that you would risk death by testing for it?  

This is the main idea of the paper. Here we show that ``quantum immortality'' is one of infinitely many rules that can be added to quantum mechanics without contradicting current scientific evidence. Many of these rules are testable by you, the reader, using a simple quantum optics experiment. 

This paper proposes a set of real experiments that can be performed by the reader to test for a unique class of new fundamental laws of physics. 


The main idea hinges on violating a fundamental law in quantum mechanics called the Born rule. The Born rule dictates that the probability of measuring a particular state is the square of that state's associated probability amplitude.

\begin{figure*}
  \includegraphics[width=1.0\linewidth]{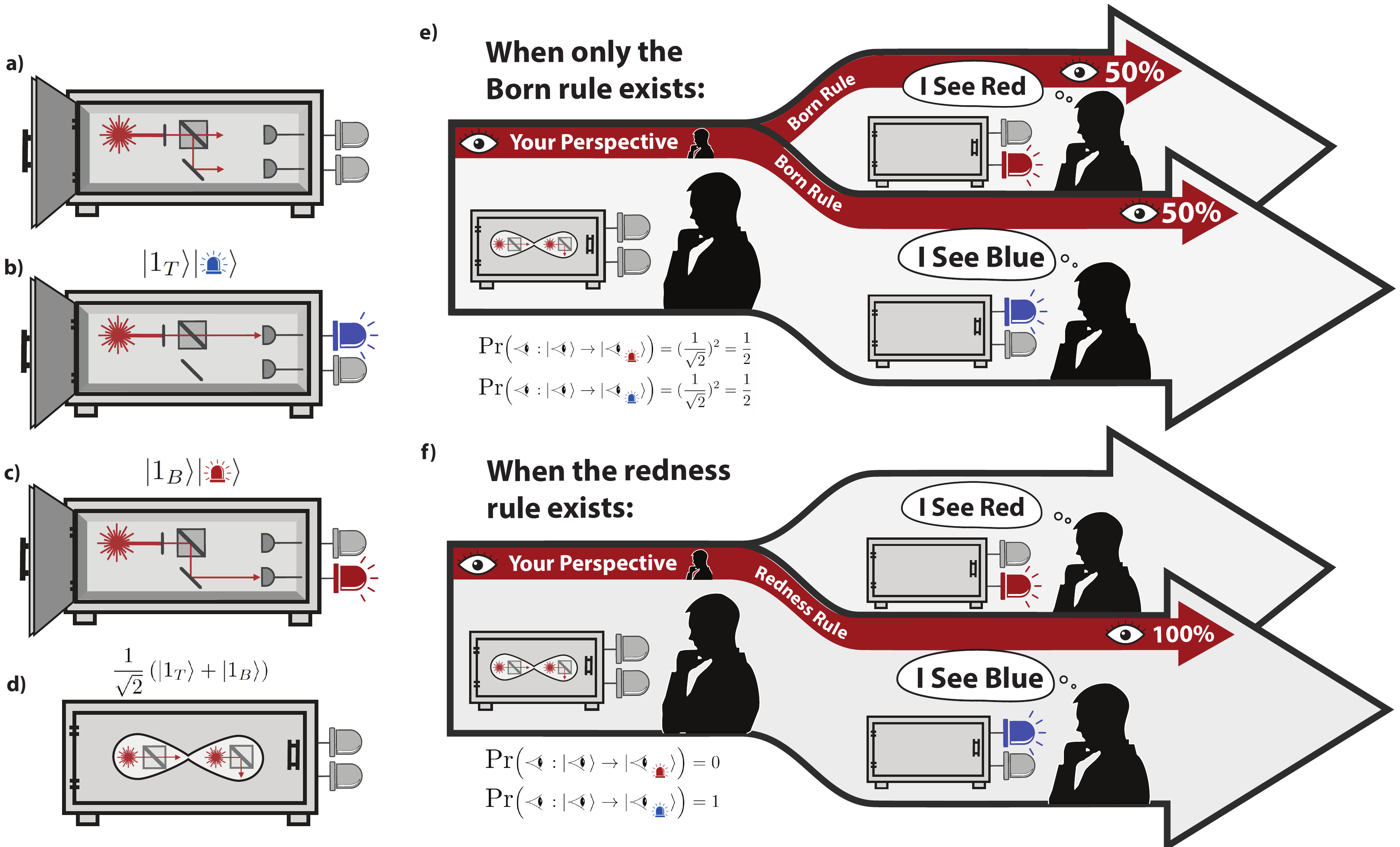}
  \caption{\textbf{Simple experiment to test for the ``Redness Rule,'' an easy to understand example of a Born-rule-violating rule.}  This figure shows the difference in outcomes in the experiment when the Redness Rule exists vs. when just the Born rule exists. This figure is paired with Figure 2, which illustrates why this experiment must be performed by you, the reader, and no one else.  a) Inside a sealable box, a single photon is prepared in an equal superposition of two paths. b) if it travels along one path, the photon hits a detector that triggers a blue light and c), traveling the other path triggers a red light. d) In the observer's perspective, when the box is sealed, the which-way information about the photon is collapsed only when the observer experiences either the red or blue lights. Ordinarily, it is expected that this observation will invoke the Born rule, and the observer will find themselves observing blue half the time and red half the time. In e) this is illustrated in the many-worlds interpretation by the red worldlines. f) If the Redness Rule exists, then in the observer's perspective, a different outcome occurs. By construction (assuming $g=1$), the observer will always find themselves in the worldline where they see blue. The difference between these outcomes of e) and f) can easily be found by repeating this experiment and performing a simple statistical analysis. As illustrated by Figure 2, such a rule is not inconsistent with our evidence of the Born rule, because if anyone else other than you, the reader, performs this test, they will not observe evidence of the Redness Rule.}
  \label{fig:RednessExperiment}
\end{figure*}

This paper is not claiming that the Born rule is false. Countless experiments have directly and indirectly confirmed that the Born rule describes our reality,\footnote{Max Born in his 1954 Nobel Lecture \cite{BornLecture} cites \cite{Born1, Born2, Born3, Born4,Born5, Born6, Born7, Born8, Born9, Born10, Born11, Born12, Born13, Born14, Born15, Born16, Born17, Born18, Born19, Born20, Born21} an extensive list of many of the original theoretical and experimental papers which developed and built on the Born rule.} and therefore it is exceedingly unlikely that the Born rule is violated in a conventional experiment.

Instead, the premise of this paper is that there may exist extra laws of quantum mechanics (we will refer to these as extra ``rules''), that can only be observed in an observer-specific reference frame. These extra rules dictate how an observer experiences wave function collapse in that observer's point of view in a way that can violate the Born rule. Explained in SM I, these rules can be interpreted as how an individual's conscious experience traverses through the worldline branches in Everett's ``many worlds'' multiverse.  

In this paper we will explain how there can exist extra rules that have the following features:

	• These rules are completely consistent with current scientific evidence of quantum mechanics.
	
	• These rules are ``observer dependent'' and consequently can only be discovered by you, the reader, if you are to perform the experiment yourself. By the nature of the rules, if anyone else performs an experiment to test for any of these rules, then in your reference frame, you will always observe that they do not observe any extra rules.

	• Many of the rules we consider are testable by you, the reader, and can be ruled out or confirmed with a simple experiment.
	
We provide an outline for four different simple quantum optics experiments that can test for the existence of four different types of Born-rule-violating rules. And (this will be repeated many times for emphasis), you, the reader, must perform these experiments to observe a meaningful result.

\section{Simple Example}

In this section we will go over a simple example of an observer-dependent rule that can easily be experimentally tested by you, the reader. Additionally, we will explain why, if anyone else other than you, the reader, were to test for this rule, evidence of its existence will not be found. 

Suppose we add the following observer-dependent rule to standard quantum mechanics, 
\begin{named}[Redness Rule: ]  Upon measurement of a quantum system, you, the reader, are less likely to experience yourself in worldlines in which you experience the sensation of redness. In cases where measurement of a quantum system is not made by your experience of redness, the standard quantum mechanical outcome occurs -- i.e., you find yourself in a worldline with a probability associated with the Born rule.  
\end{named}

Additionally a more quantitative expression of this Redness Rule can be constructed and expressed as:
\begin{align*}
&\textbf{The General Redness Rule:}\\
 &\begin{cases}
 &\scalebox{1.5}{Pr}\Big(\lateraleye : |\text{\lateraleye} \rangle \rightarrow |\text{\lateraleye}_{\detectorRed}  \rangle \Big) =  f \\
 &\scalebox{1.5}{Pr}\Big(\lateraleye : |\text{\lateraleye}  \rangle \rightarrow |\text{\lateraleye}_{\detectorBlue}  \rangle \Big) = g \\
 & \textbf{otherwise, Born Rule}\\
  \end{cases}
\end{align*}

\noindent  Overall, this represents that the observer will experience seeing redness with probability f, and experience seeing blueness with probability g. 

When $f = g = \frac{1}{2}$, the output probabilities follow the Born rule for this setup, and therefore a Born-rule-violating rule exists for $f \neq \frac{1}{2}$. There are therefore an infinite set of viable Born-rule-violating rules that can exist, each with a different value of f. 

 This Redness Rule itself is not special and is simply chosen because it is easy to understand. As discussed in more detail in SM II, the ``Redness Rule'' simply replaces the outcome of death with the experience of redness in a more general version of the no death rule. While there are some rules that have compelling motivations to exist (discussed in SM IV), we have no reason to believe that this specific rule about experiencing redness is more worth considering than another rule about experiencing greenness or a different experience entirely, apart from some kinds of rules that are unlikely, which is discussed in SM III.




Here we will explain a potential experiment that you, the reader, could perform to test if the Born rule is violated because of the Redness Rule. Additionally, we will explain how simultaneously, if \textit{someone else} were to perform this same test and tell you the results, this person would always tell you that the Born rule is not violated, and therefore the Redness Rule was not observed.

Consider the following setup: A quantum coin $|Q\rangle = \frac{1}{\sqrt{2}}\left(|H\rangle + |T\rangle\right)$ is measured by a detector that flashes a red light when the quantum coin is heads $|H\rangle$, and flashes a blue light when the coin is tails $|T\rangle$. Therefore the final state can be written as either $|H\rangle|\detectorRed\rangle$ or  $|T\rangle|\detectorBlue\rangle$, where $|\detectorRed\rangle$  and $|\detectorBlue\rangle$ represents the detector flashing red and blue respectively. Such a system can easily be experimentally performed with weak laser light and single photon detectors, as shown in Figure 1a, and outlined in SM V.

After observation, the observer either observes a flashing blue light or a flashing red light, represented by states $|\text{\lateraleye}_{\detectorRed}\rangle$ and $ |\text{\lateraleye}_{\detectorBlue} \rangle$, which denotes the observer seeing red and blue respectively. And, as discussed in the review of the Wigner's friend problem in SM I, to an external observer, the state is in a superposition of states $|H\rangle|\detectorRed\rangle |\text{\lateraleye}_{\detectorRed} \rangle$ and  $|T\rangle|\detectorBlue\rangle |\text{\lateraleye}_{\detectorBlue} \rangle$.

If you perform this experiment yourself, as shown in Figure 1f, then according to our observer-specific Redness Rule, since you are the observer, you are more likely to end up in the world $|H\rangle|\detectorBlue\rangle |\text{\lateraleye}_{\detectorBlue} \rangle$. Therefore, in your perspective the Born rule would be violated. By comparison if the rule does not exist then you are equally likely to end up in $|H\rangle|\detectorBlue\rangle |\text{\lateraleye}_{\detectorBlue} \rangle$ as $|T\rangle|\detectorRed\rangle |\text{\lateraleye}_{\detectorRed} \rangle$, as shown in Figure 1e. 

\begin{figure*}
  \includegraphics[width=1.0\linewidth]{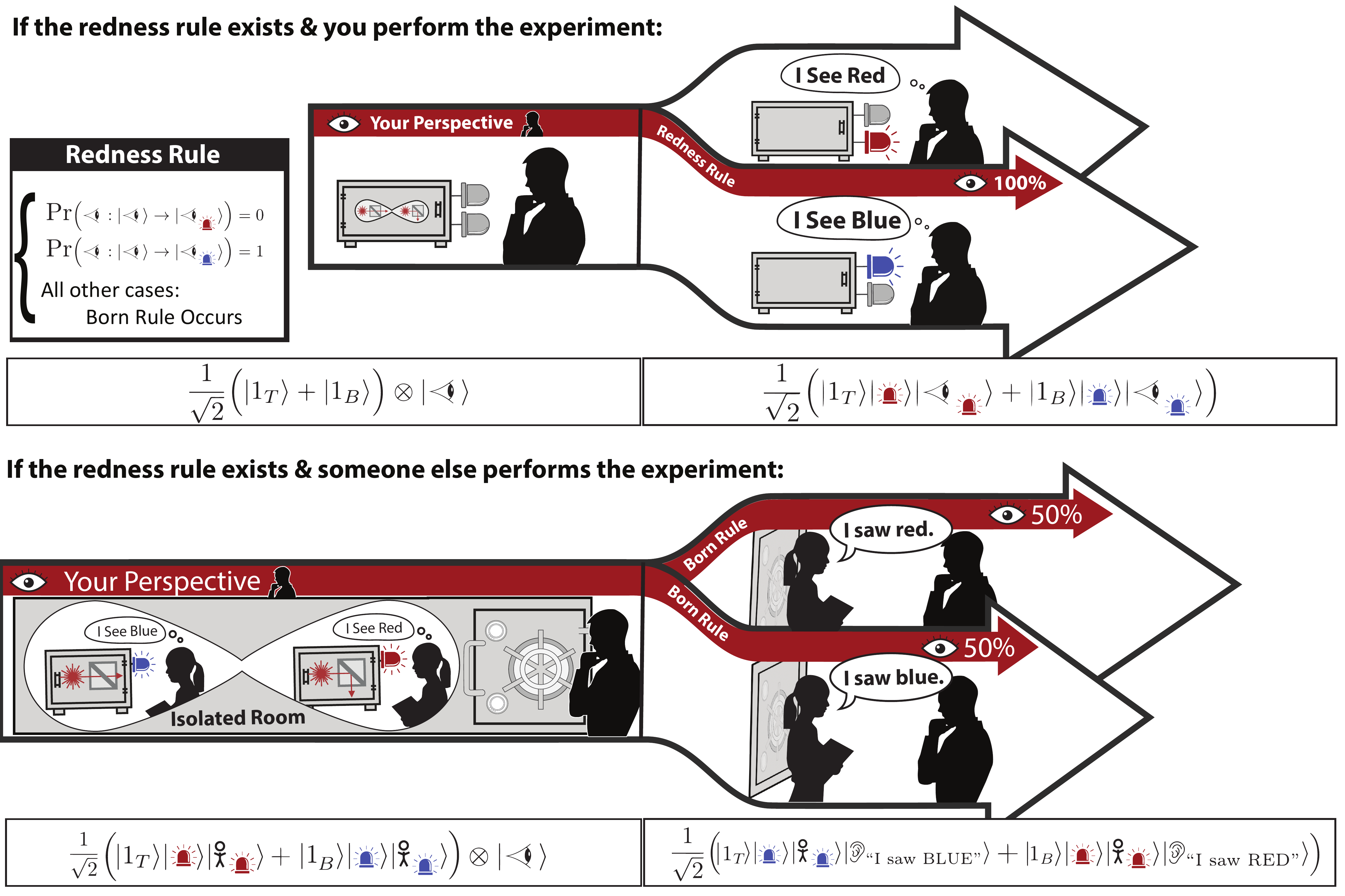}
  \caption{\textbf{Assuming the Redness Rule exists, this illustrates the difference in results when you perform the experimental test for the Redness Rule vs. when someone else performs the same experiment.} This difference is why the Redness Rule can only be tested by you, the reader, and explains how, if someone else other than you were to perform the experiment, they will never observe any evidence of the Redness Rule. The Redness Rule is meant as an arbitrary simple example to show how the existence of such an observer-dependent rule is not in contradiction with scientific evidence. Rules with more compelling motivations, described in detail in SM II, can be tested with a simple modification of the redness experiment, and we propose four similar experiments to perform, detailed in SM V. The top half of this figure illustrates what happens in the many-worlds interpretation when you, the reader, perform the test for the Redness Rule. First, before measurement, the observer is uncorrelated with his measurement. After measurement, there are two worldlines: one where the observer experiences blue and one where the observer experiences red. As shown in the black inset, if the Redness Rule exists, when your experience of the color collapses the state (assuming g=1), the special condition of Redness Rule is invoked and causes you to always experience yourself in the universe where the blue outcome has occurred. The bottom half of the figure shows how, even when the Redness Rule exists, other observers performing the experiment will not be observed \textbf{in your perspective} to have violated the Born rule.  On the bottom-left, you have someone else perform the experiment. This is a Wigner's friend situation, and as explained in SM I, until you gain information about her measurement, she is in a superposition of both outcomes. On the bottom-right, after you check the outcome by hearing her result, you find yourself in one of the two worldlines. But since you will measure the state by the experience of hearing her answer, which does not invoke the special condition of Redness Rule, you will therefore find yourself in the either worldline following the ordinary Born rule. This happens no matter what method of communication your friend communicates the outcome by, with exception to the loophole case where she communicates it by flashing a red light -- which is discussed and resolved in SM III.}
  \label{fig:Diagram1}
\end{figure*}


But, as illustrated in Figure 2, the result is different if someone else performs this experiment and reports the results to you. Consider that this other person, denoted with the symbol $ |\guy\rangle$ makes this same measurement. If this person makes a measurement when they are in a completely isolated system, then until that information can reach you, the state is still in a superposition of the different possible outcomes:
\[|\psi \rangle = \frac{1}{\sqrt{2}}\left(|H\rangle|\detectorRed\rangle |\guy_{\detectorRed} \rangle + |T\rangle|\detectorBlue\rangle |\guy_{\detectorBlue} \rangle \right)
\]
But now when you, the reader, make a ``measurement'' by asking the other person what he or she measured, the output state in your perspective is:
\begin{align*} 
|\Psi \rangle = & \\
    & \frac{1}{\sqrt{2}}|H\rangle|\detectorRed\rangle |\guy_{\detectorRed} \rangle |\ear_{\text{``I saw RED''}}\rangle \\
     + & \frac{1}{\sqrt{2}}|T\rangle|\detectorBlue\rangle |\guy_{\detectorBlue} \rangle |\ear_{\text{``I saw BLUE''}}\rangle
\end{align*}
In the first possibility, your friend sees red and she says to you ``I saw red,'' and therefore you hear her say this and that state is expressed as $|\ear_{\text{``I saw RED''}}\rangle$. Since our rule is that \textbf{``An observer, in her point of view, is less likely to experience events that makes her experience the sensation of redness, and otherwise simply experiences events following the Born rule.''} Therefore, n your perspective, by distinguishing between the two states by listening \textbf{you did not experience the sensation of redness}, and therefore the system behaves as traditionally expected, following the Born rule. So in essence, if this special observer-specific law existed, it would not appear to exist unless you, the reader, are involved in the experiment. 

\textbf{This shows that this observer-specific rule is still very much a viable theory of reality, because even if this rule existed, any experimentalist other than you cannot detect this additional rule. Therefore, until you, the reader, perform the redness experiment yourself, there is no current evidence for or against the new rule.} This is the most important point to understand in this paper. The standard methods used in forming scientific consensus cannot observe these new rules. This is because, as we have shown, as long as someone else other than you performs this measurement, they will not observe the new rule in your point of view. 

For example, we expect that if a scientist gathered thousands of people to perform this redness test, even if the Born-rule-violating Redness Rule existed, the scientist would not conclude that this new rule exists and instead would observe that the thousands of people see the red and blue outcomes according to the Born rule.

\section{General Born-rule-violating rules}
There is an infinite set of possible variants of the Redness Rule. 

Furthermore, we can imagine more similar Born-rule-violating rules if we simply switch which colors are being considered. And for that matter, the experience of the sensation of a color could be replaced with a different sensation such as sounds or sensations. But going even deeper, the actual structure of the rule can change as well.

Different types of plausible rule structures are considered in SM II. The rules we have described so far are what we categorize as ``output-at-collapse-dependent" rules, which are rules where Born-rule-violating rule triggers specifically at the time of collapse. In SM II, we explain how it is often difficult to test for this type of rule. To see a quick example of this issue: the Redness Rule triggers when the observer experiences redness, but otherwise the Born-rule occurs. Therefore, when testing the Redness Rule, if you can observe the outcome of measurement in any way before you actually collapse it by experiencing redness or blueness, the Born-rule-violating rule will not be applied -- and you would not find evidence of the Redness Rule even if it existed. 

These ``output-at-collapse-dependent" rules are one of four types of rules that we categorize in SM II, which are based on time-intervals around superposition and collapse:

\begin{itemize}
  \itemsep0em
  \item ``output-at-collapse-dependent"
  \item ``output-after-collapse-dependent''
  \item ``output-during-superposition-dependent''
  \item``output-before-superposition-dependent''
\end{itemize}

The structure of different rules are fundamentally different depending on which of the four time intervals are involved. For example, rules that are ``output-during-superposition-dependent'' are challenging to measure because of the experimental difficulty in creating superposition states over a long time interval, while ``output-after-collapse-dependent'' can be difficult to test for because of their far-reaching impact. 

\section{Forbidden Rules}
There are some nontrivial constraints that prevent any arbitrary rule from existing. In SM III, we devise a method for determining if a rule is implausible \textsl{a priori}. 

We are assuming in this paper that conventional experiments will never measure Born-rule violation, and consequently we expect that any Born-rule-violating rule which predicts that it is possible to measure Born-rule violation for external observers is implausible.  We codify this in SM III-B as the ``Scientific Consensus Meta Rule.'' But as described in SM III-D, a technical loophole stops such a strict rule about rules from working. Instead we construct a few rules of thumb to identify which rules are plausible.  

 \begin{named}[Consensus Consistency Heuristic: ] 
The more likely it is that an external observer would inadvertently invoke the rule when communicating the results of his or her measurement of the rule, the less plausible the rule is. 
\end{named}
This rule is not intuitively obvious, but it essentially arises from the assumption that you, the reader, should not expect to observe others violating the Born rule -- a result we are assuming is implausible because it would undo scientific consensus on what is already known about quantum mechanics. The rule of thumb's unintuitive construction is to deal with the mentioned loophole case, discussed in detail in SM III-E. 

Additionally, discussed in SM III-F and SM V, this rule is particularly useful for guaranteeing initial calibration of an experiment for a Born-rule-violating rule. Essentially, the Consensus Consistency Heuristic deems rules that involve typical communication as unlikely. Therefore, someone else can calibrate the experiment and communicate their results to you without fear that the communication ruins the experiment by triggering a different Born-rule-violating rule. 

In SM III, we also arrive at another important rule of thumb: 
\begin{named}[Observer Experience Heuristic: ]   Rules should depend on the experience of the observer to be plausible.
\end{named}
The intuitive explanation for this heuristic is the following: if a Born-rule-violating rule depends exclusively on physical states instead of your personal experience, then it is possible that it could be observed by an external observer, which we are assuming is implausible. This is an interesting result as it means that the only relevant Born-rule-violating rules involve your actual conscious experiences.






\section{Motivations}
Based on the heuristics mentioned in the previous section, overall a sufficiently viable rule must depend on the experiences of the observer, should not be invoked during typical methods of communication, and should not significantly contradict past experience of the observer. But other than that any arbitrary rule could exist since there is no evidence for or against it until you perform an experimental test. 

The earlier example of the ``Redness Rule" was arbitrarily chosen because it is easy to explain and has a straightforward experimental setup.  However, there is no specific reason to expect that the Redness Rule in particular would actually exist over any other viable rule. Still, the Redness Rule is useful in that it serves as a template that can be modified to test for different rules that depend on any type of experience. 

There is this vast set of possible observer-dependent rules that could exist that you, the reader, would need to test for. One potential strategy is to take an Occam's razor inspired approach and prioritize ruling out simple rules, with the rule of thumb that particularly complicated rules are probably unlikely. The first and fourth proposed experiments are designed with this in mind and can be in principle modified to cover a large set of arbitrary rules.



Alternatively, instead of searching the entire possibility space for arbitrary rules, one could look for compelling motivation for their existence. As discussed in SM IV, there are in fact some relevant motivating ideas for the existence of certain Born-rule-violating rules. These ideas are not meant to be complete, but simply to provide inspiration and spark interest -- such that it is clear that not all possible testable rules are completely arbitrary.



One motivation discussed in SM IV is that Born-rule-violating rules can potentially provide a connection between fundamental conscious experiences, sometimes called qualia, and physical processes. Furthermore, in SM IV we discuss how this could help to resolve problems relating to free will, given the existence of Born-rule-violating rules that connect your conscious experience to your physical brain. This allows for a framework in which actual qualia related to choosing a decision can steer your conscious experience to the worldline where your brain makes that decision. As discussed in greater detail in SM IV, while this could potentially resolve problems for you, the reader, it cannot answer questions about others without additional assumptions, like entanglement sharing between different brains, that are outside the scope of this paper. 

Additionally, in SM IV, we discuss an additional motivation to investigate rules based on good or bad experiences. it is possible that Born-rule-violating rules have been constructed in an altruistic manner such that particularly bad negative experiences are less likely. An agent who created the universe might have made it so that there are certain limits on things like pain, for example. It is for this reason that one of the three experiments we recommend involve receiving shocks of pain, to test for this rule.  




\section{Proposed Experiments}
In SM V, we recommend a number of different experiments that can be easily performed to get started with testing for observer-dependent rules. Here we try to cover a wide range of different possible rules from what has been discussed. 

We will provide a setup for a total of 4 experiments, testing for a number of different types of rules:
\begin{itemize}
  \itemsep0em
  \item Experiment 1: General Redness Rule
  \item Experiment 2: General Pain Rule
  \item Experiment 3: Far-reaching Impact Rules
  \item Experiment 4: Pain Steering Redness Rule 
\end{itemize}

Experiment 1 tests for the Redness Rule discussed in the main text and functions as a template for the other experiments. 

The Experiment 2 shows how the Experiment 1 can be easily modified to test for rules governing different experiences. The experience of the color red is exchanged for the experience of pain, which we are specifically motivated to investigate.


The third test and fourth experiments look for rules that have a fundamentally different rule structure. Experiment 3 looks for ``output-after-collapse-dependent” rules by trying to produce an outcome with a far-reaching impact. It does this by using the measured output results to construct a random binary sequence that will be used to purchase a real lottery ticket. Winning a real lottery is a fairly simple-to-perform method of having enough of a far-reaching impact to trigger Born-rule-violating rules that depend on high impact results.

Finally, Experiment 4 looks for ``output-before-collapse-dependent" rules, which naturally have a more complicated structure. In the example considered, the existence of pain before collapse can ``steer'' the result by changing the properties of the Redness Rule. While the rule Experiment 4 is testing for is arbitrary, it serves as a simple template experiment that can used to test for rule types that are ``output-before-collapse-dependent."

\subsection{Statistical Analysis}
The statistical analysis required to confirm if the Born rule is violated is provided in SM VI. In it we recommend ruling out a given Born-rule-violating rule by performing an experiment with enough data such that the false negative rate is below 5\%. That is, a false negative rate of 5\% means that 5\% of the time this Born-rule-violating rule will falsely be ruled out. As the Born-rule-violating rule predicts the outcome distribution is closer to our fifty-fifty Born-rule calibration, exponentially more data is required to keep a 5\% false negative rate. 

Consequently, a finite amount of measurements can only confirm a subset of rules for the class of the same rule structure with weight f. For the proposed experiments 1, 2 and 4, we recommend taking $\approx 300$ measurements, which gives a false negative rate below 5\% for Born-rule-violating rules that have weights of $f \ge 60\%$, as shown in Figure 4. For comparison, it would require over 1500 measurements to go from $f \ge 60\%$ to $f \ge 55\%$. 

The false-positive rate, when the Born rule is seen to be violated but no rule exists, is comparatively less of a problem. If a positive result ever occurs, then you simply have to repeat the experiment a second time to reduce the possibility the outcome occurred by chance. 


\section{Conclusions}
We have outlined a series of recommended tests that you the reader can perform to test for observer-dependent Born-rule-violating rules. These tests are simply recommendations and this paper serves as a first step in exploring this unexplored domain. 

There are a number of different improvements and developments that could be made. Future experiments, with the advancement of quantum memories, could explore ``output-during-superposition-dependent” rules. Additionally, since each experiment requires you the reader to manually measure each outcome yourself, this requires a large amount of data and time for experimental verification of small deviations in the Born rule. Both theoretical and experimental improvements to speed up this process would significantly allow for more rules to be tested for. 

Additionally, exploring some of the additional assumptions mentioned in the SM IV that are outside the scope of this paper is another potential promising direction. 

Finally, we mention that because the unconventional nature of the paper, we have additionally provided a section of commonly asked questions in Appendix II.  

\begin{acknowledgements}
\section{acknowledgements}
We thank Gabriel Kooreman, Ben Schweid, and Fernando Araiza Gonzalez who have been kind, encouraging and open to discussion of these ideas over the past five years. We thank Gabriel Kooreman for carefully proofreading the paper and providing some very significant feedback on several drafts of the paper, including the need for a partner to check the results of the calibration. We thank Oscar Lazo Arjona for very sharp and pointed feedback, which led to the creation of SM III to precisely address his concerns.   
Additionally, we thank Shang Yu, Aniruddha Tamma, Santi Sempere Llagostera, Jerzy Szuniewicz, Paul Burdekin, Guillaume Thekkadath, and Sonali Gera for providing helpful discussions and feedback. 
\end{acknowledgements}

\clearpage

\renewcommand\appendixpagename{Supplementary Material}
\begin{appendices}

\section{{\large SM I: Review of introductory concepts.}}
In this section we provide a short summary of the necessary prerequisite concepts for understanding the main idea of the paper. This main idea proposes a general set of observer-dependent rules which dictate how your conscious experience traverses Everett's many-worlds multiverse when you make an observation. Consequently, understanding this requires a good understanding of ``quantum immortality,'' which in turn requires a good understanding of the many-worlds interpretation of quantum mechanics. 

Because historically the many-worlds interpretation is Everett's solution to the paradox created in a thought experiment now called ``Wigner's friend,'' first we will explain what the Wigner's friend paradox is, and then we will explain how the many-worlds interpretation is a solution to that paradox. 

\subsection{Review of Wigner's Friend}

In the standard interpretation of quantum mechanics, sometimes called the ``Copenhagen interpretation,'' superpositions when they are measured ``collapse'' to become a particular outcome with a probability according to the Born rule. The ``Wigner's Friend'' thought experiment, asked by Eugene Wigner in 1961 \cite{WignersBook}, shows that this traditional interpretation of collapse leads to an apparent paradox. 

In short, if a friend of Wigner's makes a measurement of a quantum state, that friend will observe wavefunction collapse. But if Wigner places his friend inside a box that's completely isolated from the environment -- to Wigner and anyone outside the box, Wigner's friend appears to turn into a superposition of having measured both possible outcomes. Yet while those outside see her in a superposition, Wigner's friend never observes herself in a superposition of having measured both outcomes. Instead the state in her perspective is in one of two outcomes. This is a real paradox, as there is a real disagreement about what the true output state is during measurement.       

The details of the mathematics to explain this paradox is fairly straightforward. 

First consider the simple case that Wigner himself makes a measurement of a quantum state, such as: $|\psi\rangle =  c_a  |A\rangle + c_b |B\rangle$. Then, according to the Born rule, the superposition state disappears, and our state either becomes $|A\rangle$ with probability $|c_a|^2$ or it becomes $|B\rangle$ with probability $|c_b|^2$. 

But if instead Wigner has his friend perform the measurement and seals his friend in a ``closed room'' (a room completely isolated from the outside world, ensuring that there's no way Wigner can know the outcome of his friend's measurement), then the situation is different. The closed room, completely isolated from the environment is a closed system, which, if not measured by Wigner, should evolve ``unitarily'' in time. This is the main essence of the paradox: \textit{wavefunction collapse cannot occur for isolated systems, even if measurement is occuring inside that isolated system}. For example, in the Schr\"{o}dinger's cat experiment, it is the act of opening the box that changes the state. If the cat forever remained isolated, it would forever remain in a superposition. This means that even after Wigner's friend has made a measurement, as long as Wigner is isolated from his friend, no collapse can occur in Wigner's perspective. Instead, the quantum state of Wigner's friend inside the closed room is a superposition state of the form:
\[
|\psi\rangle =  \tilde{c}_a |F_A\rangle |A\rangle + \tilde{c}_b|F_B\rangle |B\rangle,
\]
where $|F_A\rangle$ (and $|F_B\rangle$) denotes the state of Wigner's friend measuring the state $|A\rangle$ (and $|B\rangle$). The proof for this is shown in Appendix
I. After his friend makes a measurement, in Wigner's perspective, the state inside the box is a quantum superposition of $|F_A\rangle |A\rangle$ \textbf{and} $|F_B\rangle |B\rangle$. Conversely, in his friend's perspective, after checking the state of $|\psi\rangle$ the friend is either in $|F_A\rangle |A\rangle$ \textbf{or} $|F_B\rangle |B\rangle$, a fundamentally different state.\footnote{it is critically important here to know that difference between a superposition and a normal probabilistic state is not simply semantic. A state $\frac{1}{\sqrt{2}}(|0\rangle+|1\rangle)$, even though there is a 50-50 chance of measuring $|0\rangle$ and $|1\rangle$ is not the same state as flipping a fair coin. When it is said that the state is in ``both'' $|0\rangle$ and $|1\rangle$, it is not simply due to our lack of knowledge about the state, but is because it can exhibit quantum interference that cannot happen with a normal flip of a coin. In our Wigner's friend situation, the friend see's himself in this normal ``coin-flip'' state (formally called a ``statistical mixture''), $\frac{1}{2}|F_A\rangle |A\rangle \langle A | \langle F_A|+\frac{1}{2}|F_B\rangle |B\rangle \langle B | \langle F_B|$. On the other hand, in Wigner's perspective, Wigner's friend is not in a statistical mixture but a superposition state. }



\subsection{Review of Many World's Interpretation}

In the Wigner's friend paradox, only when Wigner decides to open the closed room does the state collapse in his perspective, even though his friend had already made a measurement inside the closed room. This thought experiment was used by Eugene Wigner to suggest that it is consciousness within the mind that causes collapse, since Wigner cannot get the state to collapse by using his friend  --  and instead must make the measurement himself \cite{WignersMindBody}.

Wigner's student,\footnote{While Hugh Everett III published his famous doctoral dissertation at Princeton with John Wheeler as his advisor, this was after he changed his PhD thesis advisor from Eugene Wigner \cite{EverettHistory}. The first written discussion of the ``Wigner's friend problem \cite{WignersMindBody},'' which received its name due to Wigner in his publication``Remarks on the mind-body question," was actually in Everett's thesis \cite{EverettCollection}. } Hugh Everett III, had a different interpretation. If all of the possibilities in a superposition state in some sense exist, then it would explain this apparent paradox between different perspectives.

In the thought experiment, Wigner sees his friend in a superposition state of the form:
\[|\psi\rangle =  \tilde{c}_a |F_A\rangle |A\rangle + \tilde{c}_b|F_B\rangle |B\rangle.
\]

\noindent In Everett's interpretation, this is because there really is an entire universe where the state is $ |F_A\rangle |A\rangle$, and a universe where the state is $ |F_B\rangle |B\rangle$.\footnote{Historically speaking, this reference to there being different universes or ``many worlds'' was never explicitly made by Everett himself \cite{EverettPaper}. He did though propose the exact math that is used here  --  that there is only one single unitary evolution of a single wavefunction describing the entire universe, which likely implies what others later called the ``many worlds interpretation'' \cite{DeWittPopularization}.} Wigner's friend can only see \textit{either} $|F_A\rangle |A\rangle$ \textit{or} $|F_B\rangle |B\rangle$ because his measurement forces him to either be in the world with state $|F_A\rangle |A\rangle$ or the world with state $|F_B\rangle |B\rangle$. To Wigner, who has not yet checked his friend's outcome, the state is still in a superposition of the form: 
\[|W\rangle \otimes(\tilde{c}_a |F_A\rangle |A\rangle + \tilde{c}_b|F_B\rangle |B\rangle).
\]
 
\noindent Similar to when his friend makes a measurement, when Wigner asks his friend the outcome of the experiment, he forces himself to be entangled to the outcome:
\[\tilde{c}_a |F_A\rangle |A\rangle |W_A\rangle + \tilde{c}_b|F_B\rangle |B\rangle |W_B\rangle.
\]

\noindent And like his friend, inside our state there now there is a universe with a version of Wigner who confirms his friend saw $|W_A \rangle$, and a universe with a version with $|W_B \rangle$.  

Regardless of interpretation, it appears as though things when not measured are in a superposition state of possible outcomes. The only time this is not the case is when you, the reader, make a measurement of a state. But we also know that to someone isolated from you, you will appear as if you are in a superposition of having measured each possible outcome. This suggests that what collapse really is is just the experience of becoming one of these superposition possibilities. 

Furthermore, if you suppose there is an observer outside of this universe, who existed since the beginning of time, and this observer is completely cut off from the universe  --  then in that perspective, the state of the entire universe is in a superposition of all of the possible quantum events. Therefore, this interpretation can be extended to all measurements of superpositions, such that the entire universe is just a single unitary evolution of the wave-function, without any collapse at all.

To try to make these ideas more intuitively clear, consider the following thought experiment: 
 Consider an observer who is isolated from the universe since the beginning of time. Regardless of interpretation of quantum mechanics, it is true that collapse never occurs in the observer's perspective and the universe remains in a superposition of every possible outcome. This means that you the reader will, in his perspective, be in an infinite set of possible outcomes. For example, in this large superposition state there will be outcomes in which you were never born. If this outside observer decides later to take a peak at the universe, then according to the traditional interpretation of quantum mechanics, all of these superposition states (including states where you do and do not exist) all cease to exist and only one outcome remains.
 
 It could be seen to be more sensible to think that other outcomes simply do not disappear when measured, but remain in existence. By taking a peak at the universe, this observer is simply finding out which universe he ended up in, as opposed to physically causing an infinite amount of timelines and universes to disappear.

\subsubsection{Collapse in Many Worlds}

But this interpretation on its own does not explain why the Born rule occurs.  Phrasing this in our example: why is it that the friend wakes up in state $|F_A\rangle |A\rangle$ with probability $|c_a|^2$?  

Currently there is active debate \cite{BornDerviation1, BornDerviation2, BornDerviation3, BornDerviation4} if it is possible to derive the Born rule from the many worlds interpretation. 

Here we suggest that perhaps this is because the Born rule is not on its own a complete description of quantum systems for all observers.


\subsection{Quantum Immortality}

The idea of quantum immortality is already discussed main text, so here we will briefly relate it to our review of the many worlds interpretation. 

Quantum immortality is the idea that an observer \textit{in their perspective} cannot experience themselves dying. For example, consider the following Schr\"{o}dinger's cat situation. 

Suppose we have a quantum coin that is in a superposition of states heads and tails of the form $\frac{1}{\sqrt{2}}\Big(|H\rangle +  |T\rangle\Big)$, denoted as $|H\rangle$ and $|T\rangle$, and a quantum measurement is made. When the measured state is $|H\rangle$, it triggers a device that kills a cat, resulting in the state $|\catDead\rangle$ -- and if the collapsed state is $|T\rangle$ then the cat lives, represented by state $|\catAlive\rangle$. 

The cat is in a box that is completely closed off from Schr\"{o}dinger. Therefore just like the Wigner's friend paradox, the cat is in a superposition of both outcomes:
\[
\frac{1}{\sqrt{2}}\Big(|\catAlive\rangle  |H\rangle +  |\catDead\rangle |T\rangle\Big).
\]

\noindent The situation is the same as Wigner's friend, and just as before we know that there exists a universe where the cat is alive and a universe where the cat is dead. Schr\"{o}dinger by opening the box is simply checking which universe he is in.

\begin{figure}
  \includegraphics[width=\columnwidth ]{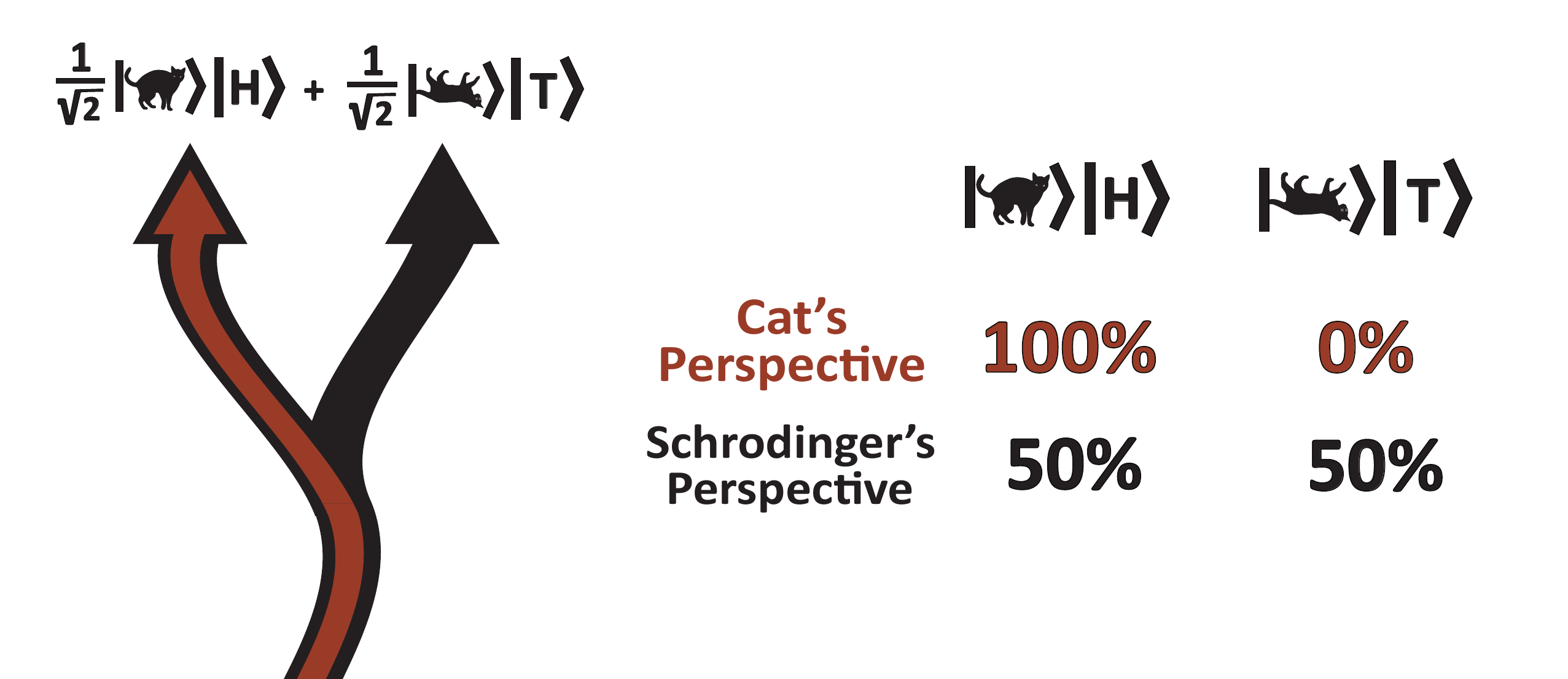}
  \caption{Simple ``multiverse diagram'' for illustrating what happens in the Schr\"{o}dinger's cat thought experiment in different perspectives when assuming the ``No Death Rule'' is correct.  The red line shows the trajectory along the many-worlds multiverse of the cat's conscious experience. Because of the the ``No Death Rule'' the probability that the cat finds itself in state $|\catAlive\rangle$ is 100\%. The black line represents the outcome in the perspective of Schr\"{o}dinger.  While the cat observes a result that violates the Born rule, Schr\"{o}dinger cannot use his results to eliminate the possibility of the existence of quantum immortality unless he instead risks death by redoing the experiment with himself performing the role of the cat.}
  \label{fig:SimpleMultiverseDiagramCat}
\end{figure}


Things are less clear for the cat. From the many-worlds interpretation, we know that there is always a universe that exists where the cat is alive. Perhaps in the conscious experience of the cat, that cat always experiences itself to be in the universe where it exists. This is the idea of ``quantum immortality.'' Observers, in their conscious experience will always find themselves in the universe where they are still observers.

There is some evidence that even Everett believed this to be true.  It is said in Eugene Shikhovtsev's biography, ``Everett firmly believed that his many-worlds theory guaranteed him immortality: his consciousness, he argued, is bound at each branching to follow whatever path does not lead to death \cite{EverettBiography}.'' 

But most of modern discussions about quantum immortality, even by believers in the many-worlds interpretation, argue that quantum immortality is incorrect. It is often argued that while it is true that there will be a version of the observer that survives, it is not rational as a decision-maker to act upon that possible reality. To quote David Deutch, ``that way of applying probabilities does not follow directly from quantum theory, as the usual one does. It requires an additional assumption, namely that when making decisions one should ignore the histories in which the decision-maker is absent....[M]y guess is that the assumption is false \cite{ImmortalityProbabilities}.'' 

Furthermore, a second common critique of quantum immortality is that this would imply that the Born-rule is violated. If this ``quantum immortality'' exists, then, in the cat's perspective in our example, the cat always lives, as illustrated in Figure \ref{fig:SimpleMultiverseDiagramCat}. Since the Born rule dictates that the probability the cat lives is $|\frac{1}{\sqrt{2}}|^2 = 50\%$, and this ``quantum immortality'' idea says the probability is 100\%, it is clear that this violates the Born rule. The Born rule is one of the most important foundational components of quantum mechanics, and quantum mechanics has countless evidence to support it. Therefore, blatantly challenging that the Born rule is not correct could be argued to be psuedo-scientific, as it is rechallenging basic knowledge that has already been discovered and established with high confidence. 




But, while ``quantum immortality'' is at odds with the standard laws of quantum mechanics, \textit{it is not in contradiction with known scientific evidence}. Quantum immortality only refers to what is observed in the point of view of the cat in the box. To Schr\"{o}dinger, or anyone else who is outside the box, the wavefunction collapses as expected (according to the Born rule) and the cat dies with $|\frac{1}{\sqrt{2}}|^2 = 50\%$ probability.

Therefore, \textit{the only way to check if quantum immortality exists is if, you the reader, put yourself in a situation like the cat in the Schr\"{o}dinger's cat thought experiment. } The experimenter must risk death to test for the existence of quantum immortality. 

In this paper we present quantum immortality as a plausible rule that could be added to quantum mechanics without contradicting current scientific evidence that the Born rule is correct. 

In the next section we will show that quantum immortality is just one of an infinite set of plausible rules that could be added to quantum mechanics that are forever unverifiable until you, the reader, perform an experiment to test for them.

\newpage
\section{{\large SM II: Observer-Dependent, Born-rule-violating Rules.}}

In this supplemental section we will define what we mean by an observer-dependent, Born-rule-violating rule. Furthermore, this section will discuss different possible rule structures, introduce a notation for defining these rules, and will categorize different types of observer-dependent rules based on how their rule structure changes their testability. 

To begin we will first reconsider quantum immortality as an extra rule that we can add to quantum mechanics. 

Following this, we will consider we will consider other types of observer-dependent rules and categorize them based on their rule structure.

\subsection{Quantum Immortality as an observer-dependent rule}

In this paper we assume that quantum immortality is not consistent with quantum mechanics. Instead, for reality to exhibit this immortality, an extra rule must be added to quantum mechanics which stipulates that observers cannot experience themselves in worldlines where they are dead. 

Here we codify ``quantum immortality'' as a rule in the following notation:
\begin{align*}
&\textbf{The No Death Rule:}\\
 &\begin{cases}
 &\scalebox{1.5}{Pr}\Big(\catAlive : |\catAlive\rangle \rightarrow |\catDead\rangle \Big) = 0 \\
 &\scalebox{1.5}{Pr}\Big(\catAlive : |\catAlive\rangle \rightarrow |\catAlive\rangle \Big) = 1 \\
 & \textbf{otherwise, Born Rule}\\
  \end{cases}
\end{align*}


\noindent Since this extra rule is outside of standard quantum mechanics, we must introduce a new notation to express it. This notation is expressing that, in the cat's perspective, the probability of it transitioning from alive to dead ($|\catAlive\rangle \rightarrow |\catDead\rangle$) is 0, the probability of transitioning from alive to alive ($|\catAlive\rangle \rightarrow |\catAlive\rangle$) is 1, and in all other situations the Born rule applies. Probability is represented by the symbol $\scalebox{1.5}{Pr}$. The observer is represented by the symbol to the left of the colon inside the parenthesis, which is $\catAlive:$ in this case. 

This notation allows us to represent quantum immortality as a sort of ``No Death Rule.''  Note that this ``No Death Rule'' is both  observer-dependent and Born-rule-violating. When Schr\"{o}dinger, represented by $\guy$ opens the box, the probability for the cat to die in our notation is: 
\[\scalebox{1.5}{Pr}\Big(\guy : |\catAlive\rangle \rightarrow |\catDead\rangle \Big).
\]
 
\noindent This does not invoke the special condition of the No Death Rule -- therefore the third ``otherwise'' case applies, and the probability Schr\"{o}dinger observes follows the Born rule. This is precisely why quantum immortality is still plausible, because it can only be confirmed if you, the reader, are the cat in a Schr\"{o}dinger's cat experiment. 

\subsection{Generalized Death Rule}
Immediately we can see that this ``No Death Rule'' can be generalized to the following rule:
\begin{align*}
&\textbf{The General Death Rule:}\\
 &\begin{cases}
 &\scalebox{1.5}{Pr}\Big(\catAlive : |\catAlive\rangle \rightarrow |\catDead\rangle \Big) = f \\
 &\scalebox{1.5}{Pr}\Big(\catAlive : |\catAlive\rangle \rightarrow |\catAlive\rangle \Big) = g \\
 & \textbf{otherwise, Born Rule}\\
  \end{cases}
\end{align*}

\noindent where f and g represent probabilities in which $f + g = 1$. Here we can see that there are an infinite set of observer-dependent, Born-rule-violating rules for the set of values $\{f\}$ for $f \in \{0, 1\}$. The key idea is that no matter the value of f, Schr\"{o}dinger, when he opens the box, still will trigger the ``otherwise'' case and simply observe the Born rule. In essence there is an infinite set of possible observer-dependent rules that could govern what happens in the perspective of the cat, which is still consistent with scientific evidence of the Born rule.   

Additionally, let us consider the same Schr\"{o}dinger's cat situation but for a general probability amplitude. That is, instead of our quantum state that we measure being $\frac{1}{\sqrt{2}}\big(|H\rangle +  |T\rangle\big)$, it is instead $c_H|H\rangle +  c_T|T\rangle$. Then with our introduced notation we can explicitly write the probabilities in this case:
\begin{align*}
    &\resizebox{.95\hsize}{!}{$\scalebox{1.5}{P}\Big(\catAlive : \Big(c_H|H\rangle+c_T|T\rangle\Big) \otimes |\catAlive\rangle \rightarrow |H\rangle \otimes |\catAlive\rangle\Big) = f$} \\
    &\resizebox{.95\hsize}{!}{$\scalebox{1.5}{P}\Big(\catAlive : \Big(c_H|H\rangle+c_T|T\rangle\Big) \otimes |\catAlive\rangle \rightarrow |T\rangle \otimes |\catDead\rangle\Big) = g$}.
\end{align*}
If f and g are constants, then it dictates that the rules for transitions $|\catAlive\rangle \rightarrow |\catDead\rangle$ occur completely independently of the underlying probability amplitudes. While this is still a possible observer-dependent rule, it seems more reasonable to consider the case where the probability amplitudes play a role in the observer-dependent probabilities. 

For that we consider rules that treat these probabilities f and g as functions: 
\begin{align*}
    &\resizebox{.95\hsize}{!}{$\scalebox{1.5}{P}\Big(\catAlive : \Big(c_H|H\rangle+c_T|T\rangle\Big) \otimes |\catAlive\rangle \rightarrow |H\rangle \otimes |\catAlive\rangle\Big) = f(c_1, c_2)$} \\
    &\resizebox{.95\hsize}{!}{$\scalebox{1.5}{P}\Big(\catAlive : \Big(c_H|H\rangle+c_T|T\rangle\Big) \otimes |\catAlive\rangle \rightarrow |T\rangle \otimes |\catDead\rangle\Big) = g(c_1, c_2)$}.
\end{align*}

\noindent where $f(c_T, c_H)$ and $g(c_T, c_H)$ are now arbitrary functions of $c_T$ and $c_H$. 

Now $f(c_T, c_H)$ and $g(c_T, c_H)$ are functions of $c_T$ and $c_H$, but must still conserve probability, and therefore $f(c_T, c_H) + g(c_T, c_H) = 1$. In the case where $f(c_T, c_H) = |c_H|^2$ and $g(c_T, c_H) = |c_T|^2$ then we have the Born rule and consequently standard quantum mechanics. The simplest modification that can be done would be to consider $f(c_T, c_H) = p|c_H|^2$ and $g(c_T, c_H) = 1 - f(c_T, c_H)= 1- p|c_H|^2$, where p is a real number from $\{0, 1 \}$. In this case, our system follows the Born rule, but with a classical probability which biases it more towards one of the two outcomes.  

In the experiments we consider, we do not try to analyze which possible functions $f(c_T, c_H)$ and $g(c_T, c_H)$ are, which is outside the scope of this paper. Instead, we assume it is sufficiently interesting identifying if a given system is violating the Born rule.


Also we note that, in an alternative notation which we will occasionally use in this paper, called ``multiverse form,'' this rule can be made more concise:


\begin{subequations}
\label{eq1}
\noindent
\begin{tikzpicture}[node distance=-.2cm and 2.5cm]
\node (A) 
    {$\big(c_H|H\rangle+c_T|T\rangle\big) \otimes |\catAlive\rangle$};
\node[above right=of A] (B) 
    {$|H\rangle \otimes |\catDead\rangle$
    };
\node[below right=of A] (C)    
    {$|T\rangle \otimes |\catAlive\rangle$};
    \draw[-{Triangle[length=3mm,width=3mm]}, line width=.65mm] (A) -- ( $ (A.0)!0.3!(B.west|-A.0) $ ) |- (B.west) node[auto,pos=0.75] {\scalebox{.8}{$f(c_T, c_H)$}};
    \draw[-{Triangle[length=3mm,width=3mm]}, line width=.65mm] (A) -- ( $ (A.0)!0.3!(C.west|-A.0) $ ) |- (C.west) node[auto,pos=0.75] {\scalebox{.8}{$g(c_T, c_H)$}} ;
\end{tikzpicture}
\end{subequations}

\noindent where each branch represents the transitions the cat experiences in the multiverse.

\subsection{Additional Similar Rules}
As discussed, the General Death Rule can only be tested by you, the reader, by being the cat in the Schr\"{o}dinger's cat experiment. To know of the existence of the Death Rule therefore requires risking death, something which is obviously too risky to reasonably test for. 

But the key point of this paper is that quantum immortality can be constructed as an extra ``observer-dependent'' rule that can be added to quantum mechanics that is not in contradiction with known scientific evidence. This means that not only can this ``Death Rule'' be generalized to the ``General Death Rule,'' but we can also easily create other observer-dependent rules with the same rule structure. 

For example, we can replace death with the experience of redness, which can be expressed in words as follows.
\begin{named}[The General Redness Rule: ]  Upon measurement of a quantum system, you, the reader, are less likely to experience yourself in worldlines in which you experience the sensation of redness. In cases where measurement of a quantum system is not made by your experience of redness, the standard quantum mechanical outcome occurs, i.e., you find yourself in a worldline with a probability associated with the Born rule.  
\end{named}

\noindent And as before this can be generalized with our notation as:
\begin{align*}
&\textbf{The General Redness Rule:}\\
 &\begin{cases}
 &\scalebox{1.5}{Pr}\Big(\lateraleye : |\text{\lateraleye} \rangle \rightarrow |\text{\lateraleye}_{\detectorRed}  \rangle \Big) =  f \\
 &\scalebox{1.5}{Pr}\Big(\lateraleye : |\text{\lateraleye}  \rangle \rightarrow |\text{\lateraleye}_{\detectorBlue}  \rangle \Big) = g \\
 & \textbf{otherwise, Born Rule}\\
  \end{cases}
\end{align*}
\noindent where $\lateraleye$ represents you, the reader. Additionally, $|\text{\lateraleye}_{\detectorRed}  \rangle$ represents you, the reader, experiencing redness, and $|\text{\lateraleye}_{\detectorBlue}  \rangle $  experiencing blueness. The variables f and g can be constants or functions of the probability amplitudes, in the same way as discussed in the General Death Rule.

This Redness Rule is the main rule that is discussed in the main text, and is illustrated in Figures 1 and 2. But, the choice to make the rule about experiencing redness is intentionally arbitrary. In principle, any state could be used to swap in instead. But this is precisely the point. Until you, the reader, begin to test for these plausible rules, they remain plausible. Though it is important to note that not all rules are plausable. We will cover later in SM III that not all rules are plausible when trying to remain consistent with scientific evidence of the Born rule. 

The next section will discuss how any rule that has the same rule structure as the Death Rule has similar difficulties in testing.


\subsection{``Output-at-collapse-dependent'' rules}

As discussed in the previous section, there is an infinite set of observer-dependent rules in which we can swap out the state of the observer dying with another output state. For all of these rules, in the perspective of the observer the probability of transitioning to a specific output state is determined by a \textit{change of state that occurs exactly at the event of collapse}. This is why we will call rules with this property ``output-at-collapse-dependent'' rules. Here we will see how these ``output-at-collapse-dependent'' rules can have a testability problem due to this rule structure. 


To see this problem, we will consider the General Death Rule as an example. So far we have stated that the only way to test for the No Death Rule is by being the cat in the Schr\"{o}dinger's cat experiment, but even this might not be enough to confirm the rule.

Let's again consider the same Schr\"{o}dinger's cat situation as considered previously. A quantum coin is prepared in the state, $\frac{1}{\sqrt{2}} \left(|H\rangle + |T\rangle\right)$. When this coin is measured by a detector, it kills the cat if the result of the detector is H and does not kill the cat if the result is T. To an external observer, the state of the system after measurement is,
\[
\frac{1}{\sqrt{2}}\Big(|\catAlive\rangle  |H\rangle +  |\catDead\rangle |T\rangle\Big),
\]
which we interpret as there being a universe where the cat is alive and a universe where the cat is dead. If you put yourself in the point of view of the cat, and the No Death Rule exists, then it appears as though in your point of view, you will always move to the state $|\catAlive\rangle  |H\rangle$ because the No Death Rule forbids you from traversing to the world with the state $|\catDead\rangle |T\rangle$. 

But there is a finite amount of time between when the device measures H and when the cat is killed. In essence, this finite amount of time it takes to kill the cat presents a problem. To an outside observer state immediately after collapse is, 
\[\frac{1}{\sqrt{2}} \left(|H\rangle |\catAlive\rangle + |T\rangle\right |\catAlive_{dying}\rangle),
\]
\noindent where $|\catAlive_{dying}\rangle$ represents the state of the cat while it is dying. If the cat can distinguish itself from being in the $|\catAlive\rangle$ state or the $|\catAlive_{dying}\rangle$ state, then it in principle can collapse the wavefunction by observing itself in the $|\catAlive_{dying}\rangle$ state. Since the rule is that \textit{only} the $|\catDead\rangle$ is forbidden and all other transitions, including $|\catAlive\rangle \rightarrow |\catAlive_{dying}\rangle$, follow the Born rule, the cat finds itself in either $|H\rangle |\catAlive\rangle$ 50\% of the time and $|T\rangle |\catAlive_{dying}\rangle$ 50\% of the time following the typical Born rule.\footnote{And, we expect the cat to perpetually remain in the $|\catAlive_{dying}\rangle$ state. Anytime there is a probability that the cat dies, the No Death Rule is invoked and the cat remains in some dying but not dead state. A scary consequence that all observers will experience if this No Death Rule is true.}

Thus, the ``quantum immortality'' rule is a difficult rule to test for, even if you are the cat in Schr\"{o}dinger's cat, because the rule depends on a change in state of the observer occurring exactly at the collapse event. 

To test for the Death Rule, it can be argued that the transition of death needs to occur instantaneously. And then, because any process causing death will have some level of continuity at some time scale, it would seem as though testing for this rule is fundamentally impossible. 

On the other hand, the event that causes the collapse in the perspective of the cat is the cat experiencing itself dying. If this process of death happens so quickly that the cat cannot actually experience itself dying, then perhaps the Born rule is not invoked. Resolving what happens in the case of the Death Rule is difficulty and consequently outside the scope of this paper. 

This difficulty is not exclusive to the Death Rule. Any rule with a similar structure to the Death Rule also has a similar problem. But for rules like the Redness Rule and the Pain Rule, where there is no confusion about what it would be like to transition to experience death, these issues are easier to resolve. 

Here we will consider an example that makes it more clear how to resolve issues with these ``output-at-collapse-dependent'' rules. Consider the following rule that makes the observer avoid worldlines in which it will experience pain:
\begin{align*}
&\textbf{The General Pain Rule:}\\
 &\begin{cases}
 &\scalebox{1.5}{P}\Big(\guy : |\guy^{\text{No Pain}} \rangle \rightarrow |\guy^{\text{Pain}} \rangle\Big) = f \\
 &\scalebox{1.5}{P}\Big(\guy : |\guy^{\text{No Pain}} \rangle \rightarrow |\guy^{\text{No Pain}} \rangle\Big) = g \\
 & \textbf{otherwise, Born Rule}\\
  \end{cases}
\end{align*}

\noindent This rule, just like the Redness Rule we constructed, simply has the consequence of death swapped out for another change in state for the observer, this time being the experience of pain.  

One way to test for this ``Pain Rule'' is to modify the experiment in the main text that tests for the Redness Rule. Instead of having the shining red and blue lights shown in Figure 1, they are replaced with a device that shocks the user. A shock of pain, or absence of a shock of pain, tells the user the which-way information of the photon and invokes the Pain Rule, if it exists. In this case the ``at-collapse-dependent'' issue is that if the collapse information leaks out to the user before they experience a shock of pain, then the Pain Rule will not be invoked. 

To see this explicitly, let's consider the following situation. An experimentalist, trying to test for this Pain Rule, purchases a device that will shock the user if the quantum state is measured to be H and will not shock them if it is T. This shocking device needs to charge before a shock is delivered and it flashes a red light when this charging process happens, and a blue light when nothing happens. In the experiment, the quantum state is measured, and if it is H, a red light flashes and then after some time the observer is shocked. This means that there will be a point in the time where the observer has distinguished between the colors but not received pain. In the external perspective, we can write this as the state: 
\[
\frac{1}{\sqrt{2}} \Big(|H\rangle|\detectorRed\rangle |NSY\rangle |\guy_\text{Red} \rangle + |T\rangle|\detectorBlue\rangle |NS\rangle |\guy_\text{Blue} \rangle \Big),
\]

\noindent where $|\guy_\text{Blue} \rangle$ represents the observer seeing blue and not yet receiving information via the shocking machine, $|\guy_\text{Red} \rangle$ represents seeing red, $|NS\rangle$ is the state of the shocking device when it is not sending a shock, and $|NSY\rangle$ (acronym for ``No Shock Yet'') is the state of the shocking-device as it is sending the shock but has not yet been sent or processed by the observer.

Now here we can see that the observer is either in the state $|\guy_\text{Blue} \rangle$ or in the state $|\guy_\text{Red} \rangle$, and therefore there is a branch where the observer sees blue and a branch which see's red. Assuming that only the Pain Rule exists, we can see that the special pain-dependent conditional is not invoked when a branching occurs and instead just the Born rule is applied.   

If instead it takes a longer period of time to observe states  $|\detectorBlue\rangle$ and $|\detectorRed\rangle$ than $|S\rangle$ and $|NS\rangle$, then the ``Pain Rule'' is testable. Our external state now becomes:
\[
\frac{1}{\sqrt{2}} \Big(|H\rangle|\text{NYR}\rangle |S\rangle |\guy^{\text{Pain}} \rangle + |T\rangle|\text{NYB}\rangle |NS\rangle |\guy^{\text{No Pain}} \rangle \Big), 
\]

\noindent where $|\text{NYR}\rangle$ and $|\text{NYB}\rangle$ represent the detector before it has been seen by the observer to be red or blue, that is ``Not Yet Red''. Therefore in this example, the key idea is that the Pain Rule can only be distinguished only if the event of pain is the event that collapses the wavefunction. If the observer can distinguish the underlying state without experiencing pain, then it makes it impossible to measure this Pain Rule.

Also, an important thing to note is that what matters is the observer \textit{experiencing} any information that can collapse the state. This is a subtle but important point that will be made more explicit in SM III's Observer Experience Heuristic. What is important for the Pain Rule is that the distinguishing information between worlds is the actual experience of pain. Additionally, the only other events that prevent this Pain Rule from triggering are other \textit{experiences} that distinguish between worlds at an earlier point in time. In SM III we will explain that this is required because it is actually impossible to come up with rules that do not depend on the experience of the observer without contradicting scientific evidence of the Born rule. But in this section, the intuition is simply that the only information that can collapse the state in the perspective of an observer is something that they can, themselves, measure. This must therefore be an actual experience and not something that can be separated from the observer. 

We suggest in this paper that an individual's conscious experience is an isolated quantum system that causes collapse only upon individual events that can be consciously experienced, often called ``qualia.'' The environment outside the observer's mind remains in a superposition, and it is always experiences of unique qualia that cause the observer to transition to different different branches of the multiverse.

Going back to our example, identifying the existence of the Pain Rule requires that the observer collapses the outcome by the experience of pain. If information leaks out to be experienced by the observer by a means other than pain, then only the Born rule will occur and the experiment is ruined. A subtle point is that it is actually not important if this information escapes out to the environment, as long as the observer cannot experience a difference in the different outcomes because of it. 


In this example testing the General Pain rule, we showed how flashing different colored lights $\{|\detectorRed\rangle, |\detectorBlue\rangle \}$ correlated with the outcome of pain could prevent testing the Pain Rule, but in principle any correlated event that can be experienced would cause an issue. One potential challenge is that the entire apparatus must have no distinguishing information that could allow the observer to identify the underlying state. This includes the device that is producing the pain. If, for example, it makes a noise or causes a non-painful tingling sensation before the event of pain, then the entire experiment is ruined.  

One potential way of combating this problem is by shortening the time interval between when the detector collapses the state and when a particular desired response is created. For example, if it only takes milliseconds from the time of collapse for the pain-device to produce pain, then it is harder for the observer to actually perceive the difference. Similarly, if the cat is killed so quickly it cannot observe itself in the $|\catAlive_{dying}\rangle$ state, then in principle the cat does not have distinguishing information before the $|\catAlive\rangle \rightarrow |\catDead\rangle$ transition. While this idea has promise, the main weakness is that it is difficult to quantify exactly how small of a time window is imperceptible and relies on the argument that the actual perceivability of the observer is a requirement of these rules. 

The best way of preventing this issue is by ensuring that the very first experience the observer will experience contains the collapse information. Methods for doing this is discussed in the proposed experiments of SM V. 

\subsection{Rules about future states}
The discussed ``at-collapse" problem stems from rules applying specifically at the event of collapse. 

An alternative solution to this issue is to simply modify the rule to fit the expected behavior. For example, is it possible that the No Death Rule can be tweaked to give the desired result when the observer is the cat in Schr\"{o}dinger's cat? By ``desired result'' we mean that the cat always ends up in the outcome $|H\rangle |\catAlive\rangle$ and never $|T\rangle |\catAlive_{dying}\rangle$. 

The easiest modification is to simply stipulate that both $|\catAlive_{dying}\rangle$ and $|\catDead\rangle$ states are forbidden. But this is not very elegant as it needs to specify what the dying state, $|\catAlive_{dying}\rangle$, is. And even when a $|\catAlive_{dying}\rangle$ state is defined, if it takes a finite amount of time to reach that state, it suffers the exact same problem as before. 

One possible rule is to suppose that output states \textit{that lead to death} are biased against. This way the $|\catAlive_{dying}\rangle$ state is Born-rule violating because it leads in the future to $|\catDead\rangle$. We call observer-dependent rules with this behavior “output-\textit{post}-collapse-dependent” conditions.

For example, consider a rule that is identical to the No Death Rule, but the state that decides the collapse is a future state at time $t_P$. We will write such a state as the following rule:
\begin{align*}
&\textbf{The After-Collapse No Death Rule:}\\
 &\begin{cases}
 &\scalebox{1.5}{Pr}\Big(\catAlive : |\overset{t_{B}}{\catAlive}\rangle \rightarrow |\overset{t_{P}}{\catDead}\rangle \Big) = 0 \\
 &\scalebox{1.5}{Pr}\Big(\catAlive : |\overset{t_{B}}{\catAlive}\rangle \rightarrow |\overset{t_{P}}{\catAlive}\rangle \Big) = 1 \\
 & \textbf{otherwise, Born Rule}\\
  \end{cases}
\end{align*}

\noindent where $|\overset{t_{P}}{\catAlive}\rangle$ represents an alive cat at time $t_{P}$. The P stands for post-collapse, time $t_{P}$ is a fixed value after $t_{@}$, the time at collapse, and $t_{B}$ is a time before collapse. In essence, the branch that the cat experiences depends on a future state that occurs after collapse. 

As before this can be completely generalized such that the probabilities are functions of the probability amplitudes:

For that we consider rules that treat these probabilities f and g as functions: 
\begin{align*}
    &\resizebox{.95\hsize}{!}{$\scalebox{1.5}{P}\Big(\catAlive : \Big(c_H|H\rangle+c_T|T\rangle\Big) \otimes |\overset{t_{B}}{\catAlive}\rangle \rightarrow |H\rangle \otimes |\overset{t_{P}}{\catAlive}\rangle\Big) = f(c_H(t), c_T(t))$} \\
    &\resizebox{.95\hsize}{!}{$\scalebox{1.5}{P}\Big(\catAlive : \Big(c_H|H\rangle+c_T|T\rangle\Big) \otimes |\overset{t_{B}}{\catAlive}\rangle \rightarrow |T\rangle \otimes |\overset{t_{P}}{\catDead}\rangle\Big) = g(c_T(t), c_H(t))$}.
\end{align*}

\noindent where $f(c_T(t), c_H(t))$ and $g(c_T(t), c_H(t))$ are now arbitrary functions of $c_T(t)$ and $c_H(t)$.



To make this more clear, here we will breifly go over what occurs at each step. 

Immediately before the measurement is made, at time $t_{B}$ (b for before-collapse), the state of the cat is not entangled and is of the form, 
\[  
\left(c_H(t_{B})|H\rangle + c_T(t_{B})|T\rangle\right) \otimes |\overset{t_{B}}{\catAlive}\rangle ,
\]
\noindent where $c_T(t_{B})$ is the probability amplitude associated with state $|T\rangle$ at time $c_T$.  As discussed previously, because of the finite time it takes to transition from $\catAlive \rightarrow \catDead$, at the time of collapse (notated $t_@$) the overall state is 
\[
c_H(t_@)|H\rangle |\catAlive\rangle + c_T(t_@) |T\rangle |\catAlive_{dying}\rangle.
\]
\noindent Finally at time $t_P$, enough time has passed for the cat to die and the state is 
\[
c_H(t_P)|H\rangle |\catAlive\rangle + c_T(t_P) |T\rangle |\catDead\rangle.
\] 
This can be written in our multiverse notation as:

\begin{widetext} 
\centering
\hspace*{-.7cm}\scalebox{1.0}{
\begin{subequations}
\label{eq1}
\noindent
\begin{tikzpicture}[node distance=-.2cm and 2.5cm]
\node (A) 
    {$\big(c_H(t_B)|H\rangle+c_T(t_B)|T\rangle\big) \otimes |\overset{t_{B}}{\catAlive}\rangle$};
\node[above right=of A] (B) 
    {$ c_H(t_@)|H\rangle \otimes |\overset{t_{@}}{\catAlive}_{dying}\rangle$
    };
\node[below right=of A] (C)    
    {$ c_T(t_@)|T\rangle \otimes |\overset{t_{@}}{\catAlive}\rangle \quad \mkern10mu  $};
\node[right=of C] (D)    
    {$ c_H(t_P)|H\rangle \otimes|\overset{t_{P}}{\catAlive}\rangle $};
\node[right=of B] (B2)    
    {$c_T(t_P)|T\rangle \otimes|\overset{t_{P}}{\catDead}\rangle$};
\coordinate (AR) at ($(A.east)-(0pt,3.2pt)$);
\coordinate (BL) at ($(B.west)-(0pt,3.2pt)$);
\coordinate (CL) at ($(C.west)-(0pt,3.2pt)$);
\coordinate (CR) at ($(C.east)-(0pt,3.2pt)$);
\coordinate (DL) at ($(D.west)-(0pt,3.2pt)$);
\coordinate (BR) at ($(B.east)-(0pt,3.2pt)$);
\coordinate (B2L) at ($(B2.west)-(0pt,3.2pt)$);
\draw[-{Triangle[length=3mm,width=3mm]}, line width=.75mm] (AR) -- ( $ (AR)!0.15!(BL|-AR) $ ) |- (BL) node[auto,pos=0.75] {\scalebox{.9}{$h(c_H, c_T)$}};
\draw[-{Triangle[length=3mm,width=3mm]}, line width=.75mm] (AR) -- ( $ (AR)!0.15!(CL|-AR) $ ) |- (CL) node[auto,pos=0.75] {\scalebox{.9}{$i(c_H, c_T)$}} ; 
\draw[-{Triangle[length=3mm,width=3mm]}, line width=.75mm] (CR) -- (DL) node[auto,pos=0.75] {} ; 
\draw[-{Triangle[length=3mm,width=3mm]}, line width=.75mm] (BR) -- (B2L) node[auto,pos=0.75] {} ; 
\end{tikzpicture}
\end{subequations}}
\end{widetext}

\noindent where $h(c_H, c_T)$ and $i(c_H, c_T)$ are the probabilities in the perspective of the cat of transitioning to the different branches. These probabilities are functions of $c_H$ and $c_T$ which are also functions of time.



But one might ask: how are future probability amplitudes used to determine probabilities for events further back in time? The key is that we can work out what these probability amplitudes are for the external observer who has not collapsed the state. In this case we first propagate the state from time $t_B$ to $t_{@}$, which represents the first time that the observer collapses the wavefunction (when the very first branch begins). Then unitary evolution of each branch can be calculated to time $t_P$, and the coefficients $c_H(t_P)$ and $c_H(t_P)$ can be determined to dictate the probabilities of being in a particular branch by the observer.  


In our example shown in the multiverse diagram, the probabilities of the cat experiencing a specific branch at time $t_@$ depends on the state of the cat at the future time $t_P$. This specific value of time $t_p$ is arbitrary so in principle all possible time values are rules that could be tested. More interestingly, a weighted sum of all possible time values could also be considered. 

For the sake of brevity, we have only considered the case where only two unique branches are formed, but in principle these rules can apply to an entire tree of branching events. For example, an after-collapse General Death Rule could be constructed to bias against branches with future states that contain death. 

The main problem that these rules have is that they can sometimes be more theoretically complicated and unwieldy. Considering rules that depend on the entire branching multiverse of all possible future states can quickly become impossible to calculate, so likely the only states that can be theoretically considered are rules in which states infinitely far away in time have infinitely small contributions. In the next section we will discuss which rules are easy and hard to test for -- and we will recommend a method to test for these more unwieldy rules. 

\section{Testing for ``after-collapse-dependent'' rules}
Some of these ``output-after-collapse-dependent'' rules can be easy to test for and some can also be very difficult to test for. 

It is often easy to see when these rules are easy to test for. For example, if the previously described experiments for the pain and Redness Rules have been ruled out, then a number of ``output-after-collapse-dependent'' rules are also simultaneously ruled out. 

To see this, consider a modification of the Redness Rule, such that the function that causes the collapse of the present state depends only on the experience of color at the future state at time $t_P$, written as:
\[
|\overset{t_{P}}{\text{\lateraleye}_{\detectorBlue}}\rangle,
\] 
\noindent where $t_P$ is some time after the time of collapse, $t_@$. This rule can be expressed in the full multiverse notation as:
\begin{widetext} 
\centering
\hspace*{-.7cm}\scalebox{1.0}{
\begin{subequations}
\label{eq1}
\noindent
\begin{tikzpicture}[node distance=-.4cm and 1.5cm]
\node (A) 
    {$  \frac{1}{\sqrt{2}}\left(|C_L\rangle|\detectorRed\rangle + |C_R\rangle|\detectorBlue\rangle \right)\otimes \overset{t_{B}}{|\text{\lateraleye}\rangle} $};
\node[above right=of A] (B) 
    {$ |C_L\rangle \otimes |\overset{t_{@}}{\text{\lateraleye}}\rangle $
    };
\node[below right=of A] (C)    
    {$ |C_R\rangle \otimes |\overset{t_{@}}{\text{\lateraleye}}\rangle $};
\node[right=of C] (D)    
    {$ |C_R\rangle \otimes|\overset{t_{P}}{\text{\lateraleye}_{\detectorBlue}}\rangle$};
\node[right=of B] (B2)    
    {$|C_L\rangle \otimes|\overset{t_{P}}{\text{\lateraleye}_{\detectorRed}}\rangle.$};
\coordinate (AR) at ($(A.east)-(0pt,3.2pt)$);
\coordinate (BL) at ($(B.west)-(0pt,3.2pt)$);
\coordinate (CL) at ($(C.west)-(0pt,3.2pt)$);
\coordinate (CR) at ($(C.east)-(0pt,3.2pt)$);
\coordinate (DL) at ($(D.west)-(0pt,3.2pt)$);
\coordinate (BR) at ($(B.east)-(0pt,3.2pt)$);
\coordinate (B2L) at ($(B2.west)-(0pt,3.2pt)$);
\draw[-{Triangle[length=3mm,width=3mm]}, line width=.75mm] (AR) -- ( $ (AR)!0.15!(BL|-AR) $ ) |- (BL) node[auto,pos=0.75] {\scalebox{1}{$h$}};
\draw[-{Triangle[length=3mm,width=3mm]}, line width=.75mm] (AR) -- ( $ (AR)!0.15!(CL|-AR) $ ) |- (CL) node[auto,pos=0.75] {\scalebox{1}{$g$}} ; 
\draw[-{Triangle[length=3mm,width=3mm]}, line width=.75mm] (CR) -- (DL) node[auto,pos=0.75] {} ; 
\draw[-{Triangle[length=3mm,width=3mm]}, line width=.75mm] (BR) -- (B2L) node[auto,pos=0.75] {} ; 
\end{tikzpicture}
\end{subequations}}
\end{widetext}

In the original redness experiment, when a specific outcome occurs, you will experience the sensation of redness and blueness for some window of time. Therefore as long as $t_P$ is inside this window, this rule is ruled out by the original experiment. In fact, this rule is actually easier to test for, as it does not have the same problem of ``output-at-collapse-dependent" rules. For the original Redness Rule, if information about which universe you are in leaks out to you before you experience redness or blueness, the experiment becomes useless. But in this modified Redness Rule, even if information leaks out to you in advance, it does not change the result of the collapse. 

Therefore, this subset of rules is in fact an easier subset of rules that are automatically obtained when the equivalent ``at collapse" versions are tested for. 

But ``output-after-collapse-dependent" rules are not all trivial, and in fact are likely the most conceptually interesting of all of the rule types. Unfortunately, these rules are very tricky to come up with a comprehensive experimental test for. The most compelling rules, as discussed in SM IV, are rules that have a large possible impact on your life. 

For example, a rule that tries to maximize your long term happiness is a very interesting rule that is very tricky to test for. In the following section, we will show a method for at trying to test for rules of this form, inspired by the well-known ``quantum suicide'' thought experiment.

\subsection{Suicide-free Quantum Suicide}
In this section we will review a well-known thought experiment connected to quantum immortality called ``quantum suicide'' -- and we will show that some interesting after-collapse-dependent rules can be tested by an experiment that resembles this ``quantum suicide'' but without the suicide. 

As discussed earlier, a difficulty of the original ``quantum immortality'' idea is that a test would require you to perform an experiment that would likely cause you to die if theory is incorrect. Additionally, as discussed in SM II the idea is actually an ``output-at-collapse-dependent" rule, which becomes difficult or impossible to test due to the possibility of collapsing the state by measuring the dying state. 

But, if these issues didn't exist, then, as explained in the original ``quantum suicide'' thought experiment, some remarkable things can be done.

The quantum suicide thought experiment is as follows. You are put in an isolated system. Outside of the isolated system, a machine makes a sequence of quantum measurements to determine a lottery ticket number and purchases a ticket associated with that number. Then after the lottery drawing is chosen, the machine kills you in the isolated system if the purchased lottery ticket is not a winning drawing. If quantum immortality exists, then in your perspective in the isolated system, you will always wake up in the universe where you survive, which is the universe where you have a winning lottery ticket. Therefore, a remarkable thing about quantum immortality is that it can be used to manipulate outcomes with fairly far-reaching consequences. 

Since the original motivation of this paper was to seek a way of testing ideas like the quantum suicide thought experiment without actually involving suicide, here we recommend a few Suicide-free Quantum-Suicide-inspired experiments. 

Since we call quantum immortality the ``No Death Rule'' in our Born-rule-violating rule framework -- the original ``quantum suicide thought experiment'' can essentially be thought of as an exploit of invoking the No Death Rule, which causes the observer to win the lottery. But in principle any Born-rule-violating rule can be invoked to assist the observer in winning the lottery, not just the `No Death Rule.' 

For example, consider a situation where the No Redness Rule exists. In this case, the same procedure can be done: a machine makes a sequence of quantum measurements to determine a lottery ticket number and purchases it; then after the lottery drawing is chosen, which we assume is chosen fairly, the machine flashes a red light to the person in the isolated system if the purchased lottery ticket is not a winning drawing. Since the No Redness Rule dictates this is impossible, in the observer's perspective the blue outcome must occur and the observer consequently will have a winning lottery drawing. 

But while this may be an interesting connection to the ``quantum suicide thought experiment'' -- and an interesting consequence of the existence of Born-rule-violating rules -- it is certainly more practical to simply use detectors flashing red and blue lights to test for the No Redness Rule. 

However, such an experiment is in fact useful for far-reaching ``output-after-collapse-dependent" rules. Consider the existence of a rule that maximizes your happiness as a rolling-average across a year. While in principle the previously described pain experiment would be able to measure a small effect, the impact of a small shock of pain across your year-long happiness is likely too small of an effect to be measured.  

However, winning the lottery is a very simple way to have a huge impact on your future in a year, as the effect will likely have far-reaching consequences. Additionally, winning the lottery is so unlikely that it serves as a reasonable way of confirming the existence of a Born-rule-violating rule without needing many measurements and careful statistical analysis. 

This is why one of the recommended experiments in SM V involve attempting to win a lottery. In this recommended experiment, you purchase a lottery ticket number based on the binary information a sequence quantum measurements of an equal superposition state. So for example, if a rule exists that maximizes your happiness as a rolling-average across a year, and winning the lottery would significantly increase your happyness, then you would have increased odds of winning the lottery.  

Now these far-reaching rules are certainly unweildy. This ``rolling-average happiness" rule is certainly not the only viable rule that could explain winning a lottery. If the lottery is won, a particularly large set of viable Born-rule-violating rules will still be plausible. Perhaps, for example, as it is often said, winning the lottery causes more long-term unhappiness than if it not won at all. Then in that case it is more likely to be evidence towards a long-term unhappiness rule rather than happiness. 

Regardless, winning the lottery in such a situation is a pretty strong indication that some sort of extra ``after-collapse" rules are in existence, but what exactly those rules are would require further experiments to rule out. Additionally, while outside the scope of this paper, additional long-term, high-impact, low-probability situations could be used to replace the act of winning the lottery -- allowing for testing of a larger span of plausible high-impact, after-collapse rules.   

Also we note that statistically speaking, someone winning the lottery is not a surprising event. Therefore it would not be surprising or interesting if you observe that some individuals have won the lottery after performing this experiment. However, if \textit{you}, the reader, win the lottery in a single run of this experiment, then there certainly would be strong evidence that an after-collapse-dependent Born-rule-violating rule exists.

\subsection{Rules about initial states}
So far we have discussed the uniqueness of ``at-collapse" and ``after-collapse" rules -- but there are still more rule types to consider. In this section we will consider ``before-collapse'' and ``during-superposition'' rules. 

Up until this point, we have considered the probability of traversing a specific branch to be a function of the probability amplitudes of states in the time interval $\{t_@, t_\infty \}$, but what about the time interval $\{-t_\infty, t_@ \}$? A number of unique types of rules can be generated using states within this new time interval. Previously we have introduced special time values $t_B$, $t_@$, and $t_p$, for times before, at, and post-collapse. Here we introduce the time, $t_{BS}$, short for before superposition, which is the time in which the state transitions to being a superposition, assuming this transition is instantaneous. Additionally we introduce the time, $t_{DS}$, short for during superposition, which is an time value contained in the interval $\{t_{BS}, t_@ \}$. Using this notation for our Schodinger's cat example, before measurement our quantum coin state is in a superposition of $|H\rangle$ and $|T\rangle$ and the combined state at time $t_{DS}$ is:  
\[
\big(c_H(t_{DS})|H\rangle+c_T(t_{DS})|T\rangle\big) \otimes |\overset{t_{DS}}{\catAlive}\rangle .
\]

\noindent There is of course a point in time, $t_{BS}$, before this superposition is created. For example if we suppose that our superposition state was originally in the state $|H\rangle$ and was unitarily rotated to the superposition state $c_H(t_{DS})|H\rangle+c_T(t_{DS})|T\rangle$, then we can write the transition of the product state in the interval from $\{ t_{BS}, t_{DS}\}$ as: 
\[|H\rangle  \otimes \overset{t_{B}}{|\catAlive}\rangle \rightarrow \big(c_H(t_{DS})|H\rangle+c_T(t_{DS})|T\rangle\big) \otimes |\overset{t_{DS}}{\catAlive}\rangle\] 
\noindent where $t_{B}$ is a time value in the interval $\{ -\infty, t_{BS}\}$. We are interested in two unique time intervals before collapse. The first is this time region in which the state is a superposition $\{ t_{BS}, t_{DS}\}$, and the second is everything that happened before the state was a superposition $\{ -\infty, t_{BS}\}$.   

Both of these time intervals are unique and different from what has been considered in the previous sections. In the previous situations, the probabilities of traversing worldline branches depends on the different outcomes that occur to the observer. For example, in our quantum immortality tests we discuss how the cat traverses different worldlines depending on the different output states of the cat. In the simplest model, it is the transition from going from alive to dead ($|\catAlive\rangle \rightarrow |\catDead\rangle$) that is forbidden.

But within each of these intervals, $\{ t_{BS}, t_{DS}\}$ and $\{ -\infty, t_{BS}\}$, no collapse is happening, the state of the cat is not becoming entangled with the measurement outcome, and therefore no worldline branching has occurred yet. This makes the rules for this region different compared to the previous sections. 

One type of rule that can exist in this time interval are something we call ``steering rules.'' These rules dictate how the branching rules change as a function of the state of the observer. 

For instance, consider a situation where a cat, before it experiences a Schr\"{o}dinger's cat situation, first can be in the state $|\catAlive_{curious}\rangle$ before any collapse happens at time $t_{BS}$. 

Now here is an example of how this state of the cat, determined before the superposition is prepared, can affect the branching rules:
\begin{align*}
&\scalebox{1.5}{P}\Big(\catAlive : |\catAlive\rangle \rightarrow |\catDead\rangle)\Big) = 
    \left\{
        \begin{array}{ll}
            1 & \quad |\overset{t_{B}}{\catAlive}\rangle = |\catAlive_{curious}\rangle \\
            0 & \quad |\overset{t_{B}}{\catAlive}\rangle \neq |\catAlive_{curious}\rangle
        \end{array}
    \right. \\
&\scalebox{1.5}{P}\Big(\catAlive : |\catAlive\rangle \rightarrow |\catAlive\rangle)\Big) = 
    \left\{
        \begin{array}{ll}
            0 & \quad |\overset{t_{B}}{\catAlive}\rangle = |\catAlive_{curious}\rangle \\
            1 & \quad |\overset{t_{B}}{\catAlive}\rangle \neq |\catAlive_{curious}\rangle
        \end{array}
    \right.\\
\end{align*}

\noindent where $t_{B}$ is a time value in the interval $\{ -\infty, t_{BS}\}$. We have defined this rule such that the probability of transitioning from $|\catAlive\rangle \rightarrow |\catDead\rangle$ is zero if the cat was curious. We call this rule ``curiosity killed the cat,'' as it is the cat's curiosity that determines if it is killed (in it is own point of view).

Both of these time intervals $\{ t_{BS}, t_{DS}\}$ and $\{ -\infty, t_{BS}\}$ can both have steering rules. 

The main difference between the two is that the during-superposition time interval $\{ t_{BS}, t_{DS}\}$ is potentially difficult to test for experimentally.  Typically, the amount of time a quantum state is in a superposition for is a very small amount of time. If, for example, a rule existed that depended on the total amount of time an observer experiences something in this state, it would not be possible to test for. 

Additionally, if a rule existed that exclusively needed an experience to occur within that time interval, this would also cause problems. For example, consider the following rule:


\begin{widetext}
\begin{align*}
&\textbf{Example Only-During-Superposition Rule:}\\
&\begin{cases}
&\scalebox{1.5}{P}\Big(\lateraleye : |\lateraleye \rangle \rightarrow |\text{\lateraleye}_{\detectorBlue} \rangle)\Big) = 
    \left\{
        \begin{array}{ll}
            1 & \quad \Big( |\overset{t_{DS}}{\lateraleye}\rangle = |\lateraleye_{\text{Pain}}\rangle \Big) \land \Big(|\overset{t_{B}}{\lateraleye}\rangle = |\lateraleye_{\text{No Pain}}\rangle \Big) \\
            \frac{1}{2} & \quad \text{Otherwise}
        \end{array}
    \right. \\
&\scalebox{1.5}{P}\Big(\lateraleye : |\lateraleye \rangle \rightarrow |\text{\lateraleye}_{\detectorRed} \rangle)\Big) = 
    \left\{
        \begin{array}{ll}
            0 & \quad \Big( |\overset{t_{DS}}{\lateraleye}\rangle = |\lateraleye_{\text{Pain}}\rangle \Big) \land \Big(|\overset{t_{B}}{\lateraleye}\rangle = |\lateraleye_{\text{No Pain}}\rangle \Big)\\
            \frac{1}{2} & \quad \text{Otherwise}
        \end{array}
    \right.\\
 & \textbf{otherwise, Born Rule}\\
\end{cases}
\end{align*}
\end{widetext}
\noindent where $|\lateraleye_{\text{Pain}}\rangle$ and $|\lateraleye_{\text{No Pain}}\rangle$ is the observer's state of feeling and not feeling pain. In this rule we have written the condition that Born-rule violation only occurs in the situation when the experience of pain only occurs during the time interval of $\{ t_{BS}, t_{DS}\}$, and nothing occurs if the experience of pain also occurs in the time interval $\{ -\infty, t_{BS}\}$. 
This ``Only-During-Superposition Pain-Steering No Redness Rule'' is laughably specific and arbitrary. This is largely in part because these ``steering rules'' essentially must have a somewhat complicated structure. While the ``before-superposition'' states are included for completion over all relevant time intervals, the ``during-superposition'' states are perhaps compelling despite the inelegance of construction of simple rules.

\subsection{Summary and Outline of many possible rules}

it is difficult to construct a list that comprehensively goes over every possible rule, but here we will do our best to try to provide a broad overview that will hopefully cover a diverse spectrum of different rule structures.

In summary of what has been discussed in this section, we first covered the way different rules can be formulated under the different time-intervals. The structure of different rules are fundamentally different depending on which of the four time intervals are involved, and we described these rules as being: 

\begin{itemize}
  \itemsep0em
  \item ``output-at-collapse-dependent"
  \item ``output-after-collapse-dependent''
  \item ``output-during-superposition-dependent''
  \item``output-before-superposition-dependent''
\end{itemize}


These rules are most often based on the state of her internal perception of the observer. It can range from low-level experiences, such as pain, colors, and sensations, to high-level experiences, such as beliefs, language, and desires.




Additionally, we have assumed the rules are constant in time, but this is also something that in principle could change as well. 

And finally, we note that all rules up until this point can in principle be combined. As long as the rule does not cause external observers to violate the Born rule, then in principle anything else is allowed.

Furthermore, this list is certainly not exhaustive, and there are likely many more interesting rules and rule strictures that can exist. 

But while there certainly is vastness in the possibility space of plausible rules, this does not stop the reader from performing experiments to narrow this down. As more experiments are performed, it should become more clear what is and is not the correct route for discovery.

\clearpage
\newpage

\newpage
\section{\large{SM III: Forbidden Born-rule-violating Rules}}

This section attempts to figure out what types of rules can exist that do not contradict our current scientific understanding of quantum mechanics. This section is long and is likely a difficult read, but it unfortunately was not clear to the author how there was a better way to present the content while keeping the ideas coherent. 
To try to make it easier on readers, we will first provide a short, complete summary of the conclusions of this section.

\section{Summary of SM III}

There are some nontrivial constraints that prevent any arbitrary rule from existing. In this section, we devise a method for determining if a rule is implausible \textsl{a priori}. 

We are assuming in this paper that conventional experiments will never measure Born-rule violation, and consequently we expect that any Born-rule-violating rule which predicts that it is possible to measure Born-rule violation for external observers is implausible.  We codify this in SM III-B as a ``meta rule,'' a rule about Born-rule violating rules, and it can be written as,
\begin{named}[Scientific Consensus Meta Rule: ] 
No rules exist that cause you, the reader, to be able to measure an external observer observing a quantum state with a probability that violates the Born rule.
\end{named}
Putting this intuitively: this means that you should not expect to see that people on their own can observe Born-rule violation, as this would simply contradict current scientific consensus on the existence of the Born rule -- and therefore all rules that conclude that in your perspective you will find other observers capable of violating the Born rule are forbidden. 

But as described in SM III-D, a technical loophole stops such a strict rule about rules from working. The problem is that you, the reader, can always theoretically observe someone else concluding that the Born rule is violated if they communicate the results to you by invoking the Born-rule violating rule. For example, if someone tells you the results of the Redness Rule, because you learned the results by hearing it, the Redness Rule is not invoked. But if someone communicates the results by flashing a red light, now you will observe that they have violated the Born rule -- in contradiction with the Scientific Consensus Meta Rule. 

While this loophole case essentially proves that the Scientific Consensus Meta Rule rules out all possible rules, we know that in reality people aren't communicating results by flashing red lights. Instead, we continue by modifying the constructed meta rules to be instead be ``rules of thumb,'' and show that this gets rid of these contradictions. 

For example, in SM III-E we work out the following rule of thumb,
 \begin{named}[Consensus Consistency Heuristic: ] 
The more likely it is that an external observer would inadvertently invoke the rule when communicating the results of her measurement of the rule, the less plausible the rule is. 
\end{named}
While its form my not suggest it, this rule of thumb replaces the Scientific Consensus Meta Rule. Instead of explicitly forbidding you from observing others observing Born rule violation, it instead forbids rules that would have been discovered by you. 

Additionally, discussed in SM III-F and SM V, this rule is particularly useful for guaranteeing initial calibration of an experiment for a Born-rule-violating rule. Essentially, the Consensus Consistency Heuristic deems rules that involve typical communication as unlikely. Therefore, someone else can calibrate the experiment and communicate their results to you without fear that the communication ruins the experiment by triggering a different Born-rule-violating rule. 

Furthermore, we also arrive at another important rule of thumb: 
\begin{named}[Observer Experience Heuristic: ]   Rules should depend on the experience of the observer to be plausible.
\end{named}
The intuitive explanation for this heuristic is the following: if a Born-rule-violating rule depends exclusively on physical states instead of your personal experience, then it is possible that it could be observed by an external observer, which we are assuming is implausible. This is a particularly interesting result as it means that the only relevant Born-rule-violating rules involve your actual conscious experiences, and the consequences of this are discussed in SM IV.

\subsection{SM III-A: Scientific Consensus Axiom}

The main purpose of the paper is to show that there's a loophole that allows for new rules to exist within quantum mechanics that are still consistent with modern evidence of quantum mechanics.

These new rules all should have the following feature:

\noindent \textbf{Any proposed new rule to be added to quantum mechanics should be undiscoverable by the scientific community through typical scientific consensus. }

This is a very important point. In this paper we are assuming that current scientific evidence sufficiently supports quantum mechanics. To go further, for our proposed new rules to be added to quantum mechanics we are going to specifically forbid any type of prediction that could be tested by the scientific community in a conventional fashion. The reason for this is that quantum mechanics has existed for around a hundred years, and it appears from almost countless different perspectives to be correct. Therefore in our new rules we propose adding to quantum mechanics, we assume that without a doubt no \textit{conventional experiment} will disprove any aspect of standard quantum mechanics.

While this assumption is straightforward, it is not immediately obvious exactly what observer-dependent Born-rule-violating rules would violate this assumption.  

As a first step to make this more explicit, we write this as an explicit assumption:

\begin{named}[Central Assumption: ] 
The observer, i.e., you the reader, will never observe that external observers conclude that the Born rule is violated.
\end{named}

Putting this in the context of the new rules we are considering, we construct the ``Scientific Consensus Axiom'':

\begin{named}[Scientific Consensus Axiom: ] 
A rule is forbidden when it causes the observer, i.e., you the reader, to observe that external observers conclude that the Born rule is violated.\footnote{Technically speaking, there is of course always a statistical chance that by chance the data looks sufficiently unlikely for the external observer to conclude that the Born rule is violated. But by ``conclude'' we assume the external observer has performed enough measurements to conclude at a sufficiently a high enough statistical confidence that the Born rule is violated.}
\end{named}

By ``external observers'' we mean any system that performs the experiment in a closed system away from you, the reader. To see mathematically what we mean by ``external observers,'' consider the Wigner's friend experiment before Wigner makes a measurement. To Wigner, who has not yet checked his friend's outcome, the state is still in a superposition of the form: 
\[|W\rangle \otimes(\tilde{c}_a |F_A\rangle |A\rangle + \tilde{c}_b|F_B\rangle |B\rangle)
\]

\noindent If you (represented as state $| \lateraleye \rangle $) were to act as Wigner in Wigner's friend experiment, then the state of the system after the friend's measurement (substituting $|W\rangle \rightarrow | \lateraleye \rangle $ and $|F\rangle \rightarrow |\guy\rangle $) is: 
\[|\Psi_O\rangle = |\text{\lateraleye}\rangle \otimes(\tilde{c}_a |\guy_A\rangle |A\rangle + \tilde{c}_b|\guy_B\rangle |B\rangle).
\]

\noindent where the ``external observer,'' state $|\guy\rangle$, performs the role of Wigner's friend. The state $|\Psi_O\rangle$ represents your perspective, when an unobserved isolated system makes a measurement of a quantum state. After you measure the outcome of the external observer the two outcome states will be $|\text{\lateraleye}_A\rangle|\guy_A\rangle |A\rangle$ or $|\text{\lateraleye}_B\rangle|\guy_B\rangle |B\rangle$. 

In standard quantum mechanics, the probability for outcome $|\text{\lateraleye}_A\rangle|\guy_A\rangle |A\rangle$ 
would be $|\tilde{c}_a |^2$ and $|\text{\lateraleye}_B\rangle|\guy_B\rangle |B\rangle$ would be $|\tilde{c}_b |^2$, following the Born rule. But of course the purpose of this paper is to suggest additional rules that cause the outcome in your perspective to not follow the Born rule. For example, there is a set of Born-rule-violating rules which do not follow the Born rule, which can be written in our notation as:
\begin{align}
    \label{eqn:thingy}
    & \scalebox{1.5}{Pr}\Big(\lateraleye : \Big(\tilde{c}_A|A\rangle + \tilde{c}_B |B\rangle\Big)  |\overset{t_{B}}{\guy}\rangle 
    \rightarrow |A\rangle|\overset{t_{@}}{\guy_A}\rangle \Big) \neq |\tilde{c}_a|^2\\
    & 
    \label{eqn:thingy2}
    \scalebox{1.5}{Pr}\Big(\lateraleye : \Big(\tilde{c}_A|A\rangle + \tilde{c}_B |B\rangle\Big)  |\overset{t_{B}}{\guy}\rangle 
    \rightarrow |B\rangle|\overset{t_{@}}{\guy_B}\rangle \Big) \neq |\tilde{c}_b|^2,
\end{align}

\noindent Here we will explain that \textbf{it is exactly this set of Born-rule-violating rules that are forbidden. } 


Our ``Scientific Consensus Axiom'' forbids rules that allow external observers to be observed concluding that the Born rule is violated. But the probabilities described in Rules \ref{eqn:thingy} and \ref{eqn:thingy2}, the output states describe the outcome when \textit{a single measurement} has been made by the external observer. A single measurement is usually not enough to conclude with some high level of statistical confidence that Born rule is violated -- but if one of the two options is sufficiently unlikely then a single measurement of the unlikely state is sufficient evidence with some level of statistical confidence that the Born rule is violated.\footnote{For example, consider if $|\tilde{c}_B|^2$ = $10^{-10}$. If in a single run the outcome is the unlikely state, Then the probability (the p-value) is $10^{-10}$. We can assume that this is a sufficiently unlikely event that we reject the null-hypothesis that the Born rule is the case. And we can therefore with a single measurement conclude that the Born rule is violated.} 

Practically speaking, this is not ideal. Just a single measurement would only be able to test for rules that violate the Born rule very substantially. 

Alternatively, the external observer can collect a set of measurements. The external observer could choose to repeat the experiment predetermined number of times until he concludes with some level of statistical confidence whether the Born rule is violated. This is more difficult to exactly quantify quantum mechanically, but an attempt is made here. 

In the case of a single measurement the external observer is in a superposition of  $|\guy_A\rangle |A\rangle$ and $|\guy_B\rangle |B\rangle$). With two measurements, we have two different sets of outcomes: states $\{|A_1\rangle, |B_1\rangle\}$ with measured outcomes $\{\guy_{A_1}, \guy_{B_1} \}$ and states $\{|A_2\rangle, |B_2\rangle \}$ with measured outcomes $\{\guy_{A_2}, \guy_{B_2} \}$. After both states are measured, the total number of measured outcomes becomes:  $$\{\guy_{A_1,A_2}, \guy_{A_1, B_2},\guy_{B_1, A_2}, \guy_{B_1, B_2} \}.$$ 

And the entire state of outcomes has the form:
\begin{align*}
    &|\tilde{c}_A|^2 |\guy_{A_1,A_2}\rangle |A_1\rangle |A_2\rangle + \tilde{c}_A \tilde{c}_B |\guy_{A_1,B_2}\rangle |A_1\rangle |B_2\rangle \\
    &+ \tilde{c}_A \tilde{c}_B  |\guy_{B_1,A_2}\rangle |B_1\rangle |A_2\rangle + |\tilde{c}_B|^2  |\guy_{B_1,B_2}\rangle |B_1\rangle |A_2\rangle
\end{align*}

As expected, no interesting quantum interference occurs between different quantum events, and the external observer simply observes an outcome with a probability associated with a two typical unfair coins with $p = \tilde{c}_A$.

For N coins, if each possible combinatoric outcome state is $|i\rangle$, and the external observer's observation of it is state $|\guy_i\rangle$, then trivially this is just a simple sum,
\begin{align*}
    &\sum_{i}^N c_i |\guy_{i}\rangle |i\rangle 
\end{align*}
In each outcome, our external observer can calculate a p-value\footnote{ For details on p-values and false positives regarding actual statistical analysis experimental data testing for Born-rule violating rules, see SM VI. } for the total outcome. Therefore our state can be reduced to:
\begin{align*}
    &\sum_{i}^N c^{V}_i |\guy^{V}_{i}\rangle |i^{V}\rangle + \sum_{j}^N c^{NV}_j |\guy^{NV}_{j}\rangle |j^{NV}\rangle 
\end{align*}

\noindent So as expected, if you the reader remain isolated from this system, this system is in superposition of the external observer concluding that the Born rule is violated and is not violated:

\[|\Psi_c\rangle = |\text{\lateraleye}\rangle \otimes \left(\sum_{i}^N c^{V}_i |\guy^{V}_{i}\rangle |i^{V}\rangle + \sum_{j}^N c^{NV}_j |\guy^{NV}_{j}\rangle |j^{NV}\rangle \right), 
\]

\noindent where $|i^{V}\rangle$ are the states in which the external observer statistically concludes the Born rule is violated, and  $|i^{NV}\rangle$ when it is not violated. Identical to the classical case, there is a probability that the Born rule is applied but that the external observer misidentifies it as being Born rule violating. This occurs with probability $ \sum_{j}^N |c^{V}_j|^2$, and is the same as the false positive classical case of determining if a coin is fair in N flips.

Our axiom essentially means that we should not expect to find $ \sum_{j}^N |c^{V}_j|^2$ is a value any different than a coin flip. That is, we should not expect to see that others (apart from you the reader) have performed the statistical analysis described in SM VI and have concluded that the Born rule is violated.

\subsection{SM III-B: Scientific Consensus Meta Rule}

Now imagine the entire scientific community is inside this closed system with the external observer. Practically speaking, this could be done if you the reader is sufficiently isolated from the external environment that the external observer and the scientific community exist in. 

Now suppose that if the external observer violates the Born rule, they write a paper about it and the results are accepted by the scientific community that the Born rule is violated. If during this period, the primary observer (you, the reader) is completely isolated from this system, then it remains in a superposition of both possibilities: 
\begin{align*}
    |\Psi_c\rangle = |\text{\lateraleye}\rangle \otimes \Big(& \sum_{i}^N c^{V}_i |\guy^{V}_{i}\rangle |i^{V}\rangle |\guy \guy \guy^{V}_i\rangle + \\
&+ \sum_{j}^N c^{NV}_j |\guy^{NV}_{j}\rangle |j^{NV}\rangle |\guy \guy \guy^{NV}_j\rangle \Big)),
\end{align*}
\noindent where the state of the outside world is in a superposition of having concluded that the Born rule is violated $|\guy \guy \guy^{V}_i\rangle$ and also not violated $|\guy \guy \guy^{NV}_j\rangle$, for each different output possibility $\{|i\rangle, |j\rangle \}$ measured by the external observer.

This is why ``scientific consensus'' is in the name of the ``Scientific Consensus Axiom.'' If an external observer can conclude a system violates the Born rule (with sufficient statistical confidence), then it implies that the experiment is conventionally testable and a typical ``scientific consensus'' can be formed.   

From this our ``Scientific Consensus Axiom'' can be made to be slightly stronger:

\begin{named}[Scientific Consensus Meta Rule: ] 
No rules exist that causes you the reader to be able to measure an external observer observing a quantum state with a probability that violates the Born rule.
\end{named}
We call this a ``meta rule,'' since it is a rule about rules. We will later show that these ``meta rules'' are too strong for their own good, as some loopholes prevent them from being useful. For now though, let us pretend these problems don't exist, as these meta rules are still instructive.

\subsubsection{Remark on Consensus Meta Rule}

Ignoring this loophole for now, the Scientific Consensus Meta Rule successfully rejects a large number of rules that would be deemed ``unscientific'' in the conventional sense. In fact, even without the loophole, this meta rule is still strong enough that very few rules are not forbidden. 

For example, consider the rule:
\begin{named}[``No Dead Cats'' Rule: ] 
In any experimentalists' perspective, when a cat is prepared in a superposition of being alive and dead, the probability of measuring a dead cat is zero. \end{named}

By definition, this rule predicts Born rule violation in the perspective of an external observer. This is in contradiction with Scientific Consensus Meta Rule, and is therefore a forbidden rule. 

Not only does this axiom rule out additional rules which directly contradict current experimental evidence, but it also rules out any conventionally testable theory that technically does not have any evidence for or against it. 

For example, consider the previous rule with a modification: when a cat is prepared in a superposition of being alive and dead \textit{on the moon}, performing a measurement on the cat will always result in the alive state. Technically speaking we do not have any scientific evidence that confirms or denies this additional rule, since technically no one has ever performed a Schr\"{o}dinger's cat experiment on the moon. But it is pretty clear that the existence of this rule is exceedingly unlikely. This ``cat on the moon'' rule is also forbidden by the Scientific Consensus Meta Rule because this rule predicts that a Born-rule-violating experiment could in principle be performed \textit{by anyone} to test it  (including external observers unrelated to you the reader).

\subsection{SM III-C: Avoiding Contradictions}
 There are some unintuitive consequences of the Scientific Consensus Meta Rule. 

In this section we will show that our Scientific Consensus Meta rule implies the following meta rule:

\begin{named}[Observer Experience Meta Rule: ] Rules must depend on the experience of the observer.
\end{named}

In an earlier example we considered the following rule:

\begin{named}[``No Dead Cats'' Rule: ] 
In any experimentalists' perspective, when a cat is prepared in a superposition of being alive and dead, the probability of measuring a dead cat is zero. \end{named}

This rule is trivially forbidden via SCMR because it explicitly says that external observers can on her own observe a Born-rule-violating rule (as observed later in your point of view, of course).

A less trivial version of the rule specifies that this is an observer-dependent rule which only applies to the point of view of you the reader:

\begin{named}[``No Dead Cats In Your Point of View'' Rule: ] 
In the point of view of you the reader, when a cat is prepared in a superposition of being alive and dead, the probability that you are in the universe with the dead cat is zero.
\end{named}

Using our notation, this rule can be written written as:
 \[
\scalebox{1.5}{P}\Big(\text{\lateraleye} : |\catAlive\rangle \rightarrow |\catDead\rangle\Big) = 0.
 \]

\noindent In other words, this rule states that in your perspective, you'll never find yourself in branches with a dead cat. 

But consider the situation where a cat is measured by an external observer. Then, according to the rule, in your perspective, you observe the following rule:
\begin{align*}
    & \scalebox{1.5}{Pr}\Big(\lateraleye : \frac{1}{\sqrt{2}}\Big(|\catAlive\rangle + |\catDead\rangle\Big)  |\overset{t_{B}}{\guy}\rangle 
    \rightarrow |\catAlive\rangle|\overset{t_{@}}{\guy_A}\rangle \Big) = 1\\
    & 
    \scalebox{1.5}{Pr}\Big(\lateraleye : \frac{1}{\sqrt{2}}\Big(|\catAlive\rangle + |\catDead\rangle\Big)  |\overset{t_{B}}{\guy}\rangle 
    \rightarrow |\catDead\rangle|\overset{t_{@}}{\guy_B}\rangle \Big) = 0,
\end{align*}

\noindent Therefore, external observers will be observed by you to have measured $|\catAlive\rangle$ in direct violation of the Scientific Consensus Meta Rule. So here we see that simply specifying that this rule applies specifically to your perspective is not enough. 

Consider the alternative rule:
\begin{named}[``No Observed Dead Cats'' Rule: ] 
In the point of view of you the reader, when a cat is prepared in a superposition of being alive and dead, the probability that you observe a dead cat is zero.
\end{named}

Using our notation, this rule can be written written as:
 \[
\scalebox{1.5}{P}\Big(\text{\lateraleye} : |\lateraleye \rangle|\catAlive\rangle \rightarrow |\lateraleye_{\scalebox{.5}{\catDead}} \rangle|\catDead\rangle\Big) = 0.
 \]
 
 \noindent which is potentially more clear in the full notation:

\begin{align*}
    & \scalebox{1.5}{Pr}\Big(\lateraleye : \frac{1}{\sqrt{2}}\Big(|\catAlive\rangle + |\catDead\rangle\Big)  |\overset{t_{B}}{\lateraleye}\rangle 
    \rightarrow |\catAlive\rangle|\overset{t_{@}}{|\lateraleye_{\scalebox{.5}{\catAlive}}}\rangle \Big) = 1\\
    & 
    \scalebox{1.5}{Pr}\Big(\lateraleye : \frac{1}{\sqrt{2}}\Big(|\catAlive\rangle + |\catDead\rangle\Big)  |\overset{t_{B}}{\lateraleye}\rangle 
    \rightarrow |\catDead\rangle|\overset{t_{@}}{|\lateraleye_{\scalebox{.5}{\catDead}}}\rangle \Big) = 0,
\end{align*}

\noindent Now, despite being almost identical to our previous rule, this rule is not forbidden. This is because this rule follows the Observer Experience Meta Rule by depending on the experience of the observer. Because the observer and the external observer are isolated from one another, it is impossible for the external observer to, on her own without the observer, perform a Born rule violating measurement (aside from a loophole mentioned later). 

This is the reasoning for the Observer Experience Meta Rule:
\begin{named}[Observer Experience Meta Rule: ] Rules must depend on the experience of the observer.
\end{named}
As seen in the “No Dead Cats In Your Point of View” Rule, as soon as a rule depends on a specific physical state that is not the observer, an external observer could in principle measure it.

But what exactly does count as an observer? Is it you specifically? For the redness test, does the observer count as the state of you when light has hit your eyes, or maybe the neurons in your brain? At some time interval before you have experienced the color red, the light has been processed by your eyes. In principle then, the state of your eyes can be thought as a physical system that can exist outside of the observer. This system then therefore could be observed by external observers and therefore would be forbidden. 

Here we suggest that, by this logic, perhaps there are no \textit{physical} states of the observer. It is the qualitative experiences of the observer, not the physical state, that counts as the observer. 

So when we refer to the state $|\lateraleye_{\detectorRed} \rangle$ , we are not referring to some specific physical process, but the conscious qualitative experience of the sensation of red. This reference to the qualitative experience is known in philosophy as ``qualia.'' The Observer Experience Meta Rule is essentially saying that the new rules that can be added to quantum mechanics must refer to your own qualia to not be inconsistent with current scientific consensus. This is a particularly unique aspect of this model and is discussed in more detail in SM IV.

\subsection{SM III-D: Issues with rigid Meta Rules}

At first, the Scientific Consensus Meta Rule appears to be a reasonable framework to describe what rules are allowed such that they do not interfere with evidence gained from typical scientific experiments. But there is a catastrophic loophole that completely destroys the viability of this meta rule. 

Because of this loophole, we will completely reject the Scientific Consensus Meta Rule. So what was the point of introducing it? Conceptually speaking, all is not lost.  We still chose to introduce these to-be-rejected Meta rules as they are conceptually helpful at understanding a more correct characterization of allowable rules. This new characterization places different possible rules on a spectrum of plausibility and is conceptually similar to the previous ideas. 

\subsubsection{Catastrophic loophole}

First we will explain this catastrophic loophole, then we will discuss how to rethink these strict meta rules into a looser spectrum of plausible rules.

Consider the most basic Redness Rule example. In it, if you the reader are the one performing the experiment, then the state before measurement is:
 \[|\text{\lateraleye}\rangle \otimes \frac{1}{\sqrt{2}}\left(|H\rangle|\detectorRed\rangle + |T\rangle|\detectorBlue\rangle \right)\]
\noindent which has the following rule:
 \[
\scalebox{1.5}{P}\Big(\text{\lateraleye} : |\text{\lateraleye}\rangle \rightarrow |\text{\lateraleye}_{\detectorRed} \rangle\Big) = 0
 \]

\noindent As discussed earlier, when you the reader make a measurement, you experience yourself in the outcome $|H\rangle|\detectorRed\rangle |\text{\lateraleye}_{\detectorRed} \rangle$ with 100\% probability, in contrast to the Born rule which would expect a probability of 50\%. 

Conversely, the opposite happens when an external observer performs the measurement. As discussed previously, when you the reader make a measurement by asking the external observer what he or she measured, the output state in your perspective is:
\begin{align*} 
    & \frac{1}{\sqrt{2}}|H\rangle|\detectorRed\rangle |\guy_{\text{RED}} \rangle |\ear_{\text{``I saw RED''}}\rangle \\
     + & \frac{1}{\sqrt{2}}|T\rangle|\detectorBlue\rangle |\guy_{\text{BLUE}} \rangle |\ear_{\text{``I saw BLUE''}}\rangle.
\end{align*}
\noindent In this situation, no longer do you experience the outcomes $\{|\text{\lateraleye}_{\detectorRed} \rangle, |\text{\lateraleye}_{\detectorBlue} \rangle \}$ but instead experience outcomes $\{ |\ear_{\text{``I saw RED''}}\rangle$,$ |\ear_{\text{``I saw BLUE''}}  \rangle \}$. Consequently, the Born-rule-violating Redness Rule is not invoked, and you observe either outcome following the Born rule. 

Therefore, it might at first seem that the Redness Rule is a viable rule, as it appears to not cause external observers to be able to be observed measuring states with non-Born rule probabilities. 

But this is actually incorrect, and there is in fact a specific method in which the Redness Rule can cause external observers to have observed Born-rule violation in her perspective. 

In this method the external observer, instead of speaking to you about his results, communicates it by flashing the same red or blue lights. Now if the Redness Rule applies, then this must be applicable in this case, and the possible total output states are then either $|H\rangle|\detectorRed\rangle |\guy_{\text{RED}} \rangle  |\text{\lateraleye}_{\detectorRed} \rangle$ or $ |T\rangle|\detectorBlue \rangle |\guy_{\text{BLUE}} \rangle |\text{\lateraleye}_{\detectorBlue} \rangle$, and therefore the outcome $ |T\rangle|\detectorBlue \rangle|  \guy_{\text{BLUE}} \rangle |\text{\lateraleye}_{\detectorBlue} \rangle$ occurs 100\% of the time. 

Therefore this predicts that the Redness Rule can cause an external observer's observations appear to violate the Born rule. But this is in violation of the Scientific Consensus Meta Rule, and therefore the Redness Rule must be rejected.

This is the catastrophic loophole that completely kills the viability of the Scientific Consensus Meta Rule. This logic can be applied to any rule. No matter what the rule is, you the reader can find yourself observing Born-rule-violating external observers if they communicate her results in a way that invokes the Born-rule-violating rule. 

Therefore if the Scientific Consensus Meta Rule exists, then it implies that there are no allowable rules. 

It might be tempting to stop here, with the argument that this loophole is proof that it is impossible for any Born-rule-violating rules to exist without directly contradicting conventional scientific evidence. But the problem with the strict language of the Meta Rule. 

The mistake with the meta rule is strictly forbidding anything that has any possibility of external observers violating the Born rule. Consider again the Redness Rule example. What are the chances that you actually already learned about general evidence about quantum mechanics through other people flashing red lights at you?  It is completely absurd to consider this a reasonable possibility in any sense. Yet it is exactly and only this circumstance that invokes the Scientific Consensus Meta Rule to forbid this rules existence. 

While the Scientific Consensus Meta Rule seems to misfire and inaccurately reject the Redness Rule, not all rules it rejects are mistakes.  

Consider again the following rule,

\begin{named}[``No Dead Cats In Your Point of View'' Rule: ] 
In the point of view of you the reader, when a cat is prepared in a superposition of being alive and dead, the probability that you are in the universe with the dead cat is zero.
\end{named}

When someone else makes measurements of these cats, you will always observe that person measuring Born rule violation, as expressed here: 
\begin{align*}
    & \scalebox{1.5}{Pr}\Big(\lateraleye : \frac{1}{\sqrt{2}}\Big(|\catAlive\rangle + |\catDead\rangle\Big)  |\overset{t_{B}}{\guy}\rangle 
    \rightarrow |\catAlive\rangle|\overset{t_{@}}{\guy_A}\rangle \Big) = 1\\
    & 
    \scalebox{1.5}{Pr}\Big(\lateraleye : \frac{1}{\sqrt{2}}\Big(|\catAlive\rangle + |\catDead\rangle\Big)  |\overset{t_{B}}{\guy}\rangle 
    \rightarrow |\catDead\rangle|\overset{t_{@}}{\guy_B}\rangle \Big) = 0,
\end{align*}
 
 \noindent In the case of the Redness Rule, there's this tiny technical case that external observers could be observed measuring Born rule violation if they signal her results to you via flashing red lights. This No Dead Cats case is very different. In this case external observers will \textit{always} be observed by you to violate the Born rule.
 
 \subsection{SM III-E: Solution via Heuristics}
 
 So we see that these two examples sit on opposite sides of a spectrum of plausibility. And now we identify that instead of our strict Scientific Consensus Meta Rule, we have a more loose rule of thumb:

 \begin{named}[Consensus Consistency Heuristic: ] 
The more likely it is that an external observer would inadvertently invoke the rule when communicating the results of her measurement of the rule, the less plausible the rule is. 
\end{named}

The Redness Rule is on the more plausible side of the spectrum, while rules based on physical outcomes are on the impossible side of the spectrum. Somewhere in the middle might be rules related to typical experiences of communication. If we considered a rule that is invoked with sounds or tones, then the possibility that an experimentalist saying to you something in the basis of: 
 
 $ \{ |\ear_{\text{``I saw RED''}}\rangle $, $|\ear_{\text{``I saw BLUE''}}\rangle   \}$
 
\noindent is higher -- and therefore via our heuristic, this rule is less plausible. 

The more likely that evidence of the rule should have already been communicated to you via standard communication, the less likely the rule. What exactly constitutes ``standard communciation'' is ambiguous, and this is why our new statement is a rule of thumb rather than a strict rule.  Going further, rules about perception of reading or hearing specific words would be even less likely to be plausible, especially as these words or ideas are closer to the typical means by which we communicate general knowledge. 

\subsubsection{Induction from past experiences}

The Consensus Consistency Heuristic is not the only rule of thumb for judging the plausibility of rules. 

Many rules can be identified as implausible by you the reader via induction. 

In general, any rule that predicts typically unlikely events to occur with high probability should naturally be ruled out by the reader if they have not already experienced these unlikely events.

For example, consider the most extreme Redness Rule which stipulates that the probability that you observe red is zero:
\begin{align*}
 &\scalebox{1.5}{Pr}\Big(\lateraleye : |\text{\lateraleye} \rangle \rightarrow |\text{\lateraleye}_{\detectorRed}  \rangle \Big) =  0 \\
 &\scalebox{1.5}{Pr}\Big(\lateraleye : |\text{\lateraleye}  \rangle \rightarrow |\text{\lateraleye}_{\detectorBlue}  \rangle \Big) = 1, \\
\end{align*}
\noindent Assuming that you aren't colorblind, then you certainly have already experienced the sensation of redness in the past, and we can assume that this rule could not have applied to you. 

On the other hand, considering the more general form of the Redness Rule:
\begin{align*}
 &\scalebox{1.5}{Pr}\Big(\lateraleye : |\text{\lateraleye} \rangle \rightarrow |\text{\lateraleye}_{\detectorRed}  \rangle \Big) = f \\
 &\scalebox{1.5}{Pr}\Big(\lateraleye : |\text{\lateraleye}  \rangle \rightarrow |\text{\lateraleye}_{\detectorBlue}  \rangle \Big) = g,
\end{align*}

\noindent this is much harder to evaluate simply from past experiences. It seems very challenging or impossible to be able to evaluate how past experiences of colors steer quantum collapse in your perspective in the general case. 

As such, we introduce a new meta heuristic:
 \begin{named}[Experience Inductive Meta Heuristic: ] 
 A rule can be deemed by you the reader as less plausible if your past experiences sufficiently provide enough evidence that contradict it.
\end{named}

Here we note an interesting remark. Could you the reader just be oblivious to these events, and even though the evidence is in front of you, you never realized it? One thing to remember is that even in your worldline, there will still be other people who can observe you. Certainly if events are particularly absurdly unlikely, then this would have likely been brought to your attention by others. 

Putting it more conceptually: while we say that rules that pass the Consensus Consistency Heuristic describe rules that external observers cannot test for themselves, this is different. Situations involving your past experiences involve you, and therefore other people observing your observations is not the same as you observing external observers. What is deemed unlikely is outside people discovering these rules on her own without you, not other people discovering these rules when you are involved. 

\subsection{SM III-F: Calibration problem}
So far we have discussed what rules are plausible \textit{before} conducting any experiments. Next, in this section we will discuss a potential difficultly in the preparation of the experiment when testing for plausible rules.

As discussed previously, in the generalization of the Redness Rule, the probability of measuring a red and blue outcome are:
\begin{align*}
 &\scalebox{1.5}{Pr}\Big(\lateraleye : |\text{\lateraleye} \rangle \rightarrow |\text{\lateraleye}_{\detectorRed}  \rangle \Big) = f(c_1, c_2) \\
 &\scalebox{1.5}{Pr}\Big(\lateraleye : |\text{\lateraleye}  \rangle \rightarrow |\text{\lateraleye}_{\detectorBlue}  \rangle \Big) = g(c_1, c_2).
\end{align*}

\noindent Violation of the Born rule occurs in the situation when $f(c_1, c_2) \neq |c_1|^2$ and $g(c_1, c_2) \neq |c_2|^2$. So for a test of the Born rule, we first prepare an experiment that has a specific value of $c_1$ and $c_2$. For simplicity let us choose $|c_1|^2 = |c_2|^2  = \frac{1}{2}$. Then we perform the experiment and see if collapsing the event via flashes of colored lights causes an outcome such that the probability of seeing red is $ \neq \frac{1}{2}$.



But there is an important detail that needs to be clarified. Here we have assumed that we have prepared a state that is in a superposition with probability amplitudes $|c_1|^2 = |c_2|^2  = \frac{1}{2}$. But how is that determined? The most obvious way is to simply collect statistics and calibrate the experiment such that the photons go into each detector with equal probability. 

But, thinking more carefully, how is that step performed? For you to do the experiment, this at some point requires you the reader to make a measurement of the outcome that formed by the statistics. A critical assumption then we have to make is that the means by which you verify this calibration does not trigger a Born-rule-violating rule. For example, what if, your monitor has a blue-light filter and white text has a reddish tint? It could be that while you thought your calibration was an unbias measurement of the experiment, you are inadvertently miscalibrating your initial state by accidentally invoking the Born-rule-violating rule. 

Easiest idea is to have a partner that does the calibration, which then tells you the results. But as we have explained in our catastrophic loophole, an external observer can still observe Born-rule violation if he or she communicates that information to you by a means that invokes Born-rule violation. 

The solution uses the idea of the Consensus Consistency Heuristic. The more a Born-rule-violating rule involves typical communication of knowledge the less plausible it exists. Therefore, for calibration of our experiment, we can have an external observer, or a set of external observers prepare the experiment. Then after preparation they will leave the isolated system and tell you the results. 

Additionally, for robustness, this can be repeated for a set of different types of typical communication. A simple example might be that the experiment is performed twice, once where the external observer prepares a written result of the preparation, and separate experiment where the calibration is done by the external observer physically tell you the result of calibration. 

\subsection{SM III-G: Combined rule problem}
So far we have clarified potential problems with calibration of the initial experiment, but are there similar problems in the measurement of actual experiment after calibration? 

There is a indeed a similar problem. There can be issues when testing for multiple individual Born-rule-violating rules that exist simultaneously.

If multiple individual Born-rule-violating rules exist simultaneously, one of the two rules can interfere with the experiment testing the other rule.



For example, the Redness Rule could exist, but if another rule \textit{also} exists, it can mess up the calibration and stop you from seeing the other rule. 

This can in principle get tricky to unravel. 

For example, consider that in addition to the Redness Rule, there also exists another rule which makes perception of a 0 less likely than a 1. Suppose, when a quantum measurement causes a single digit to appear, this rule has the form,

\begin{align*}
 \scalebox{1.5}{Pr}\Big(\lateraleye : |\text{\lateraleye} \rangle \rightarrow |\text{\lateraleye}_{\text{0}}  \rangle \Big) &= |c_0|^2 - k \\
 \scalebox{1.5}{Pr}\Big(\lateraleye : |\text{\lateraleye}  \rangle \rightarrow |\text{\lateraleye}_{\text{1}}  \rangle \Big) &=  |c_{1}|^2+k 
\end{align*}
\noindent where we assume that $0<k<c_0$. If this rule or a similar rule existed, it is possible this could interfere with the results of the redness test. For example, if every time a red or blue light flashed, under it was a 0 or a 1 indicating which outcome was selected, then this can interfere with the results. For example, if both the Redness Rule and this rule existed, its possible that it appears as though Born Rule is not violated! This would only happen in the seemingly coincidental outcome in which the effects of each rule perform an equal and opposite shift to the probabilities. So unlikely but possible nonetheless. 

Now unlike last time, it would be nonsensical to have an outside observer perform the experiment to solve this problem -- as the whole point is for you to perform the experiment yourself. 

One possible solution to the problem is to just accept that these experiments only falsify the total sets of all rules, and not individual rules.

For example, observing no Born-rule violation in the Redness Rule experiment makes a conclusion that the following statement is true: it is not possible for there to be no Born-rule-violating rules except for the Redness Rule. In other words, the test is testing for the set of rules: Born rule and no other violating rules.  

This is fairly unsatisfying, as it suggests that only the sum of all parts can be tested for, and it would be exceptionally hard to get lucky enough to guess the right set of rules to test. 

Therefore, it is desirable to go further and try to be assured that we can test for individual rules within the total set of rules. 

One potential solution for perception based experiments it to try to isolate the observer as much as possible from any possibly-correlated stimuli. This would help in the previous example, as it would prevent you from perceiving the 0 or 1 which distorts the experiment. 

But this step alone cannot on its own solve all the problems with interfering sets of rules. 

For example, suppose the Redness Rule exists and also a Pain Rule exists that tries to minimize pain. Imagine that you really want this Redness Rule to exist, but actually only this Pain Rule exists. Every time you see a blue light flash, you are happier as the odds that the Redness Rule is correct increase -- and red flashes make you more sad. In principle the Pain Rule can interfere with the redness test, and no level of isolation can prevent this.

In the end, similar to the previous conclusions, pinning down a strict set of principles for how to perfectly test for individual rules, within a set of total rules, is likely impossible. 

But, it does seem that if there really do exist sets of individual rules, the reader would have to get particularly unlucky for them to perfectly cancel out.

And if you ever do find a Born-rule-violating rule that consistently violates the Born rule, this is obviously a spectacular enough result to warrant a full investigation and do whatever is possible identify the true underlying rule that is triggering the effect.

\clearpage
\newpage

\section{{\large SM IV: Motivating Ideas}}

So far we have presented different possible Born-rule-violating rules in a very general way. This is intentional to emphasize that there is a large set of rules that are all plausible. 

Up until now this paper has been agnostic towards which rules are likely -- other than ruling out rules which end up inconsistent with scientific observations. 

Additionally, most of the example rules seem completely arbitrary. For example, the most common rule we give is the Redness Rule, which begs the question: why would any of these rules exist? Why for example would the universe manipulate quantum probabilities in my perspective such that I see red things less often? 

So far we have provided no motivating ideas for why any rules might exist. This is intentional, as part of the point is that a large motivation is just that this is a large domain of possibility that can be explored. A better understanding for why such seemingly arbitrary rules exist might come after your discovery of them. 

However, there are relevant motivating ideas for the existence of some rules. These ideas are not meant to be complete, but simply to provide inspiration and spark interest -- such that it is clear that not all possible testable rules are completely arbitrary.

\subsection{Consciousness's influence on quantum mechanics}
First we note that no matter how arbitrary, any observer-dependent rule if it were to exist would imply that conscious experiences can affect the outcome of quantum systems. 

First we note this consciousness-quantum connection does not require any special quantum properties of the brain, such as the brain being a quantum computer. As we have shown in SM III, these rules, if any are to exist, must inherently depend on the conscious experience of the observer, or otherwise be inconsistent with our assumptions of scientific consensus. This means that, if any of these rules were to exist, it would create a causal connection between the actual conscious experience of an observer and their quantum measurement. 



For example, consider that the no Redness Rule exists. In this case, your conscious \textit{experience} of the sensation of redness is what prevents certain quantum states from occurring upon measurement. Even if the light reaches your eye and enters your brain, if for whatever reason that redness is not \textit{experienced}, then this rule will not be invoked. Therefore it is the conscious experience of redness that causes the outcome that you observe. 


Perhaps it is obvious to the reader that extra rules added to quantum mechanics that must depend on the conscious experience of the observer would, if they exist, establish a connection between consciousness and quantum measurements. The main point of mentioning this is that even an arbitrary-looking rule has very interesting consequences if it were to exist. But in the next couple of sections we will explain how the existence of some Born-rule-violating rules have some interesting maybe less obvious consequences. 

\subsection{Viability for Dualism}

The consciousness-quantum connection of these Born-rule-violating rules give room for dualistic solutions to the mind-body problem that are typically rejected. 



In what is described as the ``hard problem of consciousness,'' it is not known how qualia, the name for internal conscious experiences, such as the sensation of pain or redness, are mapped to physical processes. Even if the entire physical process of the brain is understood, it is believed by many that this would be insufficient to explain or know the connection between the two. 

Qualia is often thought to be casually unconnected to the physical mechanisms that govern the brain's response, a view known as epiphenomenalism \cite{Chalmers}. This is an alternative to the more traditional dualistic model that qualia and physical processes are causally connected from both directions \cite{Dualism}. A commonly stated weakness of dualistic models is that a viable connection between qualia and physical processes seem incompatible with our models of physics \cite{Incompatible}.  

The Born-rule-violating rules could provide an alternative model, reviving a possible dualistic model of consciousness, through a potential bridge between the qualia experienced by the mind and the body governed by quantum mechanics. The potential solution would be that there are physical Born-rule-violating rules that causally connect conscious experiences to physical states, which we will describe in detail in the next section.





\subsection{Motivation via Free Will}
While a complete discussion is outside the scope of this paper, we mention that the mentioned philosophical problems are connected with the idea of free will. 

So far we have limited ourselves to Born-rule-violating rules in which certain experiences change the probabilities of ending up in different universes. But in principle it is possible to construct a completely dualistic model in which your choices are actually part of the Born-rule-violating rules. 

For example, imagine that you are in a situation where you are experiencing pain. Suppose you are tired while running in a race, and you are considering stopping. You are presented with a choice to stop running or to continue. Perhaps your physical brain actually exists in a superposition of your brain picking both choices, but which outcome you \textit{experience} is actually decided by you via a Born-rule-violating rule.


To clarify, let us consider an example construction of such an idea. Suppose there exists a blackbox function, completely unconnected to the known laws of physics, which receives your qualia (conscious experiences of you, the reader) as inputs and returns the output we will call the ``agency signal.'' Next, let us assume that a choice made by your brain between options A or B involves the collapse of a quantum state $|\psi \rangle = |A \rangle + |B \rangle$ and outcome ``A'' results in the brain state $|O_A \rangle$ and ``B'' results in $|O_B\rangle$, which leads to performing choice A and B. Then that would mean to an isolated external observer, your brain is in a superposition\footnote{This superposition should not be confused with the superposition necessary for the brain to be a quantum computer. For this example to work it is not required that the brain operates like a quantum computer, but simply that the outcome of a choice needs to be a result of a quantum measurement. In principle this is guaranteed since all classical information can be broken down into quantum information. But the idea is somewhat testable in the sense that if a brain's choice between option A or B only has an absurdly small quantum uncertainty, then it would require a Born-rule-violating rule that almost completely ignores probability amplitudes.} of $\frac{1}{\sqrt{2}}(|O_A \rangle + |O_B\rangle)$, and consequently there's a universe where your brain chooses option A and one with option B. 

Now consider the following Born-rule-violating rule that interacts with this blackbox function:

\begin{named}[The Free-Will Rule: ] When the brain of you, the reader, chooses between options A or B, which results in your brain being in a superposition of having made both choices (in the reference frame of an external observer), then the outcome you, the reader, will consciously experience is determined by the agency signal.    
\end{named}

As an analogy, the output of this blackbox function can be thought of as similar to a controller for a video game. In a video game, a player can provide inputs to the game through a controller. The game itself is programmed in a deterministic way to respond to whatever is sent via the controller, but the system of the player and the controller is indeterministic in the sense that the player, which is outside the deterministic programming of the game, can choose what inputs to send to the controller. In our analogy, the game itself is the universe, the output of the game that you see are your qualia, the inputs you send through your controller are your choices, and the player is you.     

This is a physical model in which the conscious experiences you observe cause the actual physical output you observe to be different. But it does so in a way that is not inconsistent with our functional understanding of how the brain works, apart from assuming that the brain can exist in a superposition of choices. 
Unlike the previous examples, it is not obvious what is a good test for a Born-rule-violating rule of this form, but that does not mean that a test is impossible. Future work could investigate this further to see if there is a way of identifying the existence of such a ``free-will'' Born-rule-violating rule. Related tests are discussed in SM V. 

Furthermore, the presented idea, even if the Free-Will Rule exists, does not completely resolve the free will problem as it only explains why you, the reader, have free will -- and does not explain free choices of others. Additional rules would have to be added to explain how you experience other people's choices, and it would need to be explained how experiencing other people's choices does not violate assumptions about scientific consensus. 

While outside the scope of this paper, there are potential future experiments that could be performed to investigate this further, which is discussed at the end of SM V.  

\subsection{Motivation for Rules Governing Positive and Negative Experiences}
Additionally, we present another motivating idea for the existence of Born-rule-violating rules which suggests that there may be rules governing positive and negative experiences. 

If humanity were to try create a new universe where other conscious beings could exist, one could imagine a desire in the design of that universe for a mechanism that prevents individuals from experiencing particularly bad experiences. For example, perhaps these humans would desire to create rules in this new universe to prevent conscious observers from experiencing torture. From this we can speculate that it is plausible that this universe was constructed in a similar fashion. 

It could be that such rules already exist in this universe in the form of Born-rule-violating rules. 

We know that we can observe individuals who experience dramatic negative emotions like torture, but if something like a Pain Rule existed, it would reduce the likelihood that you, yourself, ever end up experiencing such events. For example, if a Pain Rule like this exists, you will still observe others experience events like torture, but you will never end up in a universe where you, yourself, get tortured. 

The possible rules also aren't necessarily limited to avoiding negative experiments. For example a rule could exist that simply maximizes long-term fulfillment. 

This motivation is the reason for experiments 2 and 3, which are described in SM V.

\clearpage
\newpage 

\section{{\large SM V: Outline of Recommended Experiments}}

Here we suggest a number of different experiments that can be easily performed to get started with testing for observer-dependent rules. These experiments are meant to provide a template that can be used to test for a large set of rules with different structures. 

We will provide a setup for a total of four experiments, testing for a number of different types of rules:
\begin{itemize}
  \itemsep0em
  \item Experiment 1: General Redness Rule
  \item Experiment 2: General Pain Rule
  \item Experiment 3: Far-reaching Impact Rules
  \item Experiment 4: Pain Steering Redness Rule 
\end{itemize}

The core experimental setup of these experiments can be thought of as a simple quantum random number generator that is not concerned with security, but is instead focused on having the collapse of the quantum event occur by you having a specific experience. In essence, the experiments we will consider use this same core equipment and analysis, with slight changes for the particular rule being tested. For that reason, we will go into the most detail with the Redness Rule, then for other experiments will just explain the minor changes.


\subsection{Experiment 1: General Redness Rule}
In this section we will explain how an experiment can be performed to test for the Redness Rule. This experiment in particular will provide a simple template that can be used to test for a large set of rules. For example, experiments 2 and 4 are both simple modifications of this Redness Rule. For this reason, this first experiment will be covered in the most detail. Here we will outline a recommended experimental setup, explain how it tests for the Redness Rule, discuss important technical details of the experiment, and outline an experimental procedure that can be performed that implements these technical details.

\subsubsection{Experimental setup}
The recommended experimental setup is illustrated in Figure 1. Attenuated light from a  commercial laser splits single photons into two paths. We recommend using a polarizing beam splitter and wave plates to ensure the paths are evenly distributed. These paths are sent into two different single photon detectors. As described later in more detail, the photon rate that is sent into the detectors is significantly higher than the dark count rate, and a randomly selected outcome is chosen at a rate of 1 Hz. Finally, this entire system is completely isolated, with exception to the output signal. This output signal then triggers a red or blue light depending on which path the detected photon traveled. A full procedure of the measurements of the experiment will be performed when taking measurements, described in a later section. This includes the fact that you, the reader, must perform the physical experiment of identifying colors to verify any Born-rule violating properties. Additionally, the calibration of the beam splitter should be performed by a separate observer than you, the reader. 

\subsection{Review of Theory for Experiment}
While the actual experiment itself is trivially simple compared to modern quantum optics experiments, to reach a more general audience we briefly go over the basic theory for why the described setup works. 

In this section we will briefly show some elementary math explaining how the measurement of a commercial laser can be used to put the reader in the state: 
\[
\frac{1}{\sqrt{2}}\left(|\detectorRed\rangle |\text{\lateraleye}_{\detectorRed} \rangle + |\detectorBlue\rangle |\text{\lateraleye}_{\detectorBlue} \rangle \right), 
\]
which, in an external observer's perspective, is a superposition of having experienced red and blue colors. 

The objective of this first experiment is to test for the Born-rule violating properties of the generalized Redness Rule, which has the form: 
\begin{align*}
&\textbf{The General Redness Rule:}\\
 &\begin{cases}
 &\scalebox{1.5}{Pr}\Big(\lateraleye : |\text{\lateraleye} \rangle \rightarrow |\text{\lateraleye}_{\detectorRed}  \rangle \Big) =  f \\
 &\scalebox{1.5}{Pr}\Big(\lateraleye : |\text{\lateraleye}  \rangle \rightarrow |\text{\lateraleye}_{\detectorBlue}  \rangle \Big) = g \\
 & \textbf{otherwise, Born Rule}\\
  \end{cases}
\end{align*}

\noindent As described in the main text and in greater detail in SM II, this means we want a quantum state to be collapsed by either experiencing a red or a blue color. This can be performed in a number of different methods, but here we recommend a simple quantum optics experiment which works by collapsing the which-way path information of a single photon.

 In the setup of Figure 1, a commercial laser source produces a beam of light in a coherent state $| \alpha \rangle$. This beam is attenuated with an ND filter until it is well below the single photon level $|\alpha| \ll 1$, which means the attenuated state can be approximated as $\approx |0\rangle - \epsilon |1\rangle$. 
 
 This output will be measured with two single photon detectors, placed at two different paths. Since these detectors only measure single photons, the state $|0\rangle$ cannot be observed, and our measurement of these coherent states can be thought of as a post-selected measurement of $|1\rangle$, the single photon state. 
 
 While in Figure 1 we show an ideal beam splitter splitting the photon into two paths, since real beam splitters do not distribute exactly $50
\%$ of the power in each direction, for the real experiment we recommend using a polarizing beam splitter to split the light in a 50\% distribution, which will be calibrated by a separate person as specified in SM II. 

 The laser light is rotated by a sequence of waveplates to produce diagonally polarized light. Therefore the post-selected photon after these waveplates can be expressed as a superposition of horizontally and vertically polarized light: 
\[
\frac{1}{\sqrt{2}}\left(|1_H\rangle |0_V\rangle + |0_H\rangle|1_V\rangle\right),
\] 
\noindent with the subscript representing modes in polarization H and V.


This light is then sent through an ideal polarizing-beam splitter which has been tuned to split the power evenly, and therefore the state of the photon becomes an equal superposition of two paths (notated as L and R):
\[
\frac{1}{\sqrt{2}}\left(|1_L\rangle+|1_R\rangle\right).
\] 
\noindent Next, a detector is placed in the left and right paths, collapsing the state by measuring which path the photon went through. But, to an outside observer completely isolated from this system, the detectors are in a superposition state of having measured both outcomes:
\[
\frac{1}{\sqrt{2}}\left(|C_L\rangle + |C_R\rangle \right),
\] 
\noindent where $|C_L\rangle$ and $|C_R\rangle$ represents the left detector and right detector obtaining clicks respectively. 

Next, in the experiment, if the left detector clicks, a red light is flashed outside of the isolated room. And if the right detector clicks, a blue light is flashed. Therefore before the light reaches your eye and you experience either a red or blue flash, the state of the system is:
 \[|\text{\lateraleye}\rangle \otimes \frac{1}{\sqrt{2}}\left(|C_L\rangle|\detectorRed\rangle + |C_R\rangle|\detectorBlue\rangle \right).\] 
\noindent After you process the experience of color, this causes the state to transition to:  
  \[ \frac{1}{\sqrt{2}}\left(|C_L\rangle|\detectorRed\rangle |\text{\lateraleye}_{\detectorRed} \rangle + |C_R\rangle|\detectorBlue\rangle |\text{\lateraleye}_{\detectorBlue} \rangle \right).\]
\noindent This is exactly the desired superposition state that would potentially trigger our Born-rule violating Redness Rule. Therfore, if we can perform the following procedure we will be making a direct measurement of the Redness Rule.

\subsection{Experimental Isolation}

As discussed a previous section, the Redness Rule is in the class of observer-dependent rules that we defined as ``output-at-collapse-dependent'' conditions. We described that the most important component when testing for these rules is that the observer cannot observe the collapse of the quantum system other than by the specific output that is Born-rule violating. It is critical that \textbf{collapse of the experiment must occur through perception of the red and blue lights and not through some other means}. 

Therefore in the design of the experiment the person perceiving the different colored lights should be isolated as much as possible from the outcome of the experiment. If you can observe the outcome of the collapse by another means, then the Redness Rule will not be invoked -- and any Born-rule violation will not be observed, even if it exists. 

Probably the easiest method is to have the observation of the colored lights in a separate isolated room from the photons and detectors. Ideally this isolated room has no electronics or equipment which is connected to the experiment. Additionally, no other distinguishing information should be shown to you other than the colors of the lights. 

\subsection{Avoiding Dark Counts}
Because you, the reader, will be recording your observations in real time, the rate at which the light flashes red or blue needs to be slow enough for you to reliably record your results. Consequently, we propose performing a measurement via flashing LEDs once a second. 

In a perfect world with accurate detectors with zero dark counts, we would be able to do this photon-by-photon. But typical single photon detectors typically have dark counts around a hundred per second. 

For this experiment, a simple method is performed to get around this problem. To avoid errors due to dark counts, we will send photons into the detectors at a rate much larger than the dark counts -- but also at a rate small enough that the detector is not saturated and the probability of measuring pairs $|2\rangle$ is small. For example, this could be done with a typical avalanche photodiode, with a photon generation rate of about $10^5$ Hz. Then using a slow electronic trigger, a single photon from this large stream of measurements is selected to trigger the red and blue lights. Since the chance that one of these photons is from the dark counts is very small, the effects of these dark counts changing the outcome can be ignored. 





\subsection{Procedure}
Here we provide a recommended procedure for carrying out this experiment which combines all of the discussed details of the various supplemental sections.

\subsubsection{Step 1: Predetermine Number of Measurements}


User predetermines confidence interval and number of measurements. We recommend choosing $p = .05$, which implies a five percent chance of incorrectly concluding the Born rule was violated. Additionally we recommend taking 300 measurements, which gives a false negative rate below 5\% for $f > .6$, as shown in Figure 4. More details on the statistics are provided in SM VI. 

	
\subsubsection{Step 2: Calibration}
As discussed in SM III-F, the experiment can be safely calibrated by another person. A calibration of the experiment is performed by a partner in an isolated system. 
The partner tunes the angle of the half-wave plate until the average counts of both arms are equal. The partner then records the total counts over several seconds and records the ratio between the two counts. The partner then communicates the results to you, the reader. Ideally, this information is communicated to you by ``conventional means." We recommend both showing the results in a written and spoken form.

\subsubsection{Step 3: Perform the experiment}
You, the reader, will observe flashes of red and blue lights and keep a tally of the total number of red and blue flashes. This tally is the measurement of the Born rule that will be used for the analysis. 

\subsubsection{Step 4: Analysis}
Analysis of the tally of red and blue flashes is performed. A simple p-test is made to evaluate if the null hypothesis is rejected at the predetermined confidence interval. If the p-value for the data is below .05, then the data is determined to be Born-rule violating. This is discussed in detail in SM VI. 

\subsubsection{Step 5: Repeat}
Here we mention that it is possible to repeat the experiment to test for different variations or parameters. But we note that to note that the possibly of inadvertently ``p-hacking'' the results increases as the number of tests increases. Therefore, it is recommended that any significant result is then retested a second time to reduce the possibility of false positives. Further details discussed in SM VI.


\subsection{Experiment 2: General Pain Rule}
The first experiment on the redness rule serves as a general template that can test for Born-rule violating rules. Here we show how this experiment can be modified to test for a more motivationally interesting rule. Specifically we will replace the sensation of color with the experience of pain. 


Pain is something that is difficult to characterize well in a typical scientific experiment, as an experimentalist has no way of actually experiencing someone else's pain. But it is possible in this case for you to explore sensations of pain because you the reader are performing the experiment, and you have direct access to your own experience of pain. 


This second experiment will test for the pain rule described in SM II, which can be written as:
\begin{align*}
&\textbf{The General Pain Rule:}\\
 &\begin{cases}
 &\scalebox{1.5}{P}\Big(\guy : |\guy^{\text{No Pain}} \rangle \rightarrow |\guy^{\text{Pain}} \rangle\Big) = f \\
 &\scalebox{1.5}{P}\Big(\guy : |\guy^{\text{No Pain}} \rangle \rightarrow |\guy^{\text{No Pain}} \rangle\Big) = g \\
 & \textbf{otherwise, Born Rule}\\
  \end{cases}
\end{align*}
\noindent for some value of $f + g = 1$ in which $f \neq \frac{1}{2}$.

\subsubsection{Experimental Setup:}

The experiment will be the same as the Redness Rule test, with the difference that instead of having flashing red and blue lights, a device is prepared to shock the reader if one of the two outcomes occur. The reader will either experience or not experience a shock in a given time interval. Then, the reader will record the amount of shocks they received in the time interval, and this will be used to analyze if the Born rule was violated. 

As discussed previously, to properly test for this rule, it is important that the event of wavefunction collapse occurs specifically due to the experience of pain and not another means. Therefore, as discussed in SM II, it is important that the device that causes the shock of pain does not trigger any perceivable information to you before the shock is administered. 

Additionally, we mention that in our rule we do not specify exactly what $ |\guy^{\text{Pain}}\rangle$ constitutes. In principle, this rule could exist for any emotion or any degree of emotion. So there is conceptual interest even if the degree of pain is a small shock. But we do point out that, as discussed in SM IV, the potential existence of an ``anti-torture'' rule is one of the primary motivations for this test -- it may be that some of these rules remain untestable due to a limit on the amount of temporary pain that can be ethically administered.

\subsection{Experiment 3: far-reaching, high-impact rules}
While experiments 1, 2, and 4 are essentially the same experiment with slight modifications, this third experiment is fundamentally of a different form. Here we will briefly explain this third experiment, which looks for evidence of far-reaching, high-impact rules. 

As discussed in detail in SM II, both the Redness Rule and the Pain Rule of experiments 1 and 2 are both ``output-at-collapse-dependent'' conditions. There are a total of four types of rules based on specific timings of collapse:

\begin{itemize}
  \itemsep0em
  \item ``output-at-collapse-dependent"
  \item ``output-after-collapse-dependent''
  \item ``output-during-superposition-dependent''
  \item``output-before-superposition-dependent''
\end{itemize}

This third experiment will investigate ``output-after-collapse-dependent'' rules, the final experiment will test for ``during-superposition'' rule types, and the final ``before-superposition'' rule type will be discussed at the end. 

The motivation for this third experiment is discussed in detail in SM II. In short, there are some interesting, nontrivial rules that are ``output-after-collapse-dependent.'' These rules by their nature of depending on future states end up being more challenging to explicitly test for. Instead of performing a direct test for a specific rule, here we suggest performing an experiment that would give evidence that some kind of far-reaching, high-impact rule exists. 

\subsubsection{Experimental Setup}
Described in SM II, the experimental setup for this third experiment is essentially the same thing as the quantum suicide experiment, without the suicide part.

The proposed experiment is as follows: you put yourself in an isolated system sufficiently disconnected with the environment. Then, an experimentalist (or a computer) uses the experiment used for redness test to generate a random binary sequence. This quantum random number generation is used to generate a lottery ticket number. Then, after the drawing of the lottery, this lottery ticket is then verified if it is a winning ticket. Only after that are you are notified from inside the isolated system with the results.

Also, we note that since ``output-after-collapse-dependent'' generally do not suffer from problems of requiring an isolated experiment, it may not be necessary for you to be in a completely isolated system while the procedure is taking place. One thing that is certainly important is that the lottery drawing is truely quantum random, else classical determinacy could prevent the rule from being invoked. 

In essence, if you the reader perform this experiment and find that you have won the lottery, the unlikeliness of that event should suggest to you that there is potentially a  ``output-after-collapse-dependent'' Born-rule violating rule that has a form that is far-reaching and high-impact. More effort would have to be done by you to investigate what exactly this rule could be. 




\subsection{Experiment 4: Pain-Steering Redness Rule}
So far we have described experiments for two of the four types of rules based on specific timings of collapse, here we will pride a test for a third rule type: ``output-before-collapse-dependent." 

As described in SM II, ``output-before-collapse-dependent" rules can function by what we describe as ``steering rules.'' Here is an example of a simple steering rule:
\begin{align*}
&\scalebox{1.5}{P}\Big(\lateraleye : |\lateraleye \rangle \rightarrow |\text{\lateraleye}_{\detectorBlue} \rangle)\Big) = 
    \left\{
        \begin{array}{ll}
            1 & \quad |\overset{t_{BS}}{\lateraleye}\rangle = |\lateraleye_{\text{Pain}}\rangle \\
            0 & \quad |\overset{t_{BS}}{\lateraleye}\rangle \neq |\lateraleye_{\text{No Pain}}\rangle
        \end{array}
    \right. \\
&\scalebox{1.5}{P}\Big(\lateraleye : |\lateraleye \rangle \rightarrow |\text{\lateraleye}_{\detectorRed} \rangle)\Big) = 
    \left\{
        \begin{array}{ll}
            0 & \quad |\overset{t_{BS}}{\lateraleye}\rangle = |\lateraleye_{\text{Pain}}\rangle \\
            1 & \quad |\overset{t_{BS}}{\lateraleye}\rangle \neq |\lateraleye_{\text{No Pain}}\rangle
        \end{array}
    \right.\\
\end{align*}

\noindent According to this rule, the existence of pain before collapse can ``steer'' the result by changing the properties of the Pain Rule.  When you are experiencing pain, quantum events will collapse such that you always experience blueness. And conversely, when you are not experiencing pain, quantum events will collapse such that you always experience redness.

This final experiment can easily be done simply by combining the first and second experiment. Now, a shock of pain will be given to observers before they experience either color, and the experiment will otherwise continue. 

We note that this final experiment is a fairly arbitrary rule that is included for completeness -- to complete the set of rule types -- and can perhaps be omitted. After this the only rule type that has does not have an experiment are rules that are exclusively ``during-superposition'' dependent, which will be mentioned in our discussion of future experiments.

\subsection{Potential Future Experiments}
So far we have provided 4 recommended experiments that can be used as templates to test for different rule types. Here we will discuss a number of different future experiments that could also be performed.

\subsubsection{``output-during-superposition-dependent'' rules}

We have recommended performing experiments on 3 of the 4 rule types. The reason why this fourth rule type is omitted is because non-trivial rules of this form require states to be in superposition states for a measurable amount of time. Fundamentally, quantum superposition is the essence of quantum mechanics, so if we are to consider extra rules to be added that interact with quantum mechanics - considering time intervals that have to do with superposition states seems very relevant. For example, imagine a Pain Rule that steers the result depending on the amount of time the state is in a superposition. 

But there is a problem with this time interval. The amount of time a photon remains in a superposition state is very small relative to the time-scale at which perception and experience occurs. If a rule specifically requires an experience to only be within this time interval, then this time interval would need to be larger than the minimum amount of time a human can perceive something. The minimum unit of time a human can perceive is around 100 ms, while typical quantum superpositions do not last this long. 

One type of future experiment could be done to perform these types of quantum tests in the regime where quantum events exist for a perceivable amount of time. For the implementation of the experiment we consider, measuring these rule types would require a quantum memory that can store superposition states for more at least 100 ms. Future quantum memories with large storage times could be used to investigate these rule types. 

In addition to an experiment which uses future quantum memory technology, there are a number of additional experiments that could be performed in the future. 

\subsubsection{More Complicated Experiences}

In these experiments, we have limited ourselves to providing rules to a very basic set of experiences: pain and color. But in principle any of these experiences could be repeated for more complex or abstract experiences. So in principle these experiments could be repeated for different stimuli. For example, pictures of happy or sad cats could be flashed instead of colors. Nostalgic photos could be used to generate a sensation of happiness instead of pain, or outrageous pictures could be used to generate the experience of anger.

Ideally, a set of experiments could be constructed to try to span as much different types of rules and experiments as possible. In principle, experiences could be broken down into fundamental components, and tests could be performed for each possible individual sensation. So for example, the Redness Rule could be replaced with a ``smell rule,'' a ``touch rule,'' a ``taste rule,'' etc.  

While this could be done for some individual basic experiences, there's an infinite set of more complicated  and abstract experiences. Intuitively it seems impossible to test for all possible combinations of all of these different properties. But it seems reasonable that these rules in general can be falsified by covering a large span of different possible rules. 

\subsubsection{Correlations between individuals}
In SM IV, we mention possible dualistic motivations for certain Born-rule violating rules. While the existence of some rules could solve many questions related to the Mind-Body problem, a complete answer would likely need to have a connection that connects experiences of you with the experiences of others. A potential avenue to explore is to look at correlations between individuals. If for example, entanglement is shared between individual's brains, this could be a way to make further progress in completely answering these questions. 

\subsubsection{Rules about agency}
As discussed in SM IV, rule can be constructed in a dualistic manner which allow for a type of ``free will.'' It is not immediately obvious how to come up with an experiment that shows how a brain's outcome could be steered by a Born-rule violating rule -- but it is also not obvious that an experiment is not possible. 

One potential direction, other than investigating correlations between observers and quantum effects in the brain, is to look at rules related to human agency. So far we have discussed rules surrounding experiences that you the reader observe as inputs, but rules can also extend to your conscious choices that you produce as outputs. 

For example, perhaps a rule exists that increases the likelihood for the outcome that you consciously wish for. Such a test can easily be performed by modifying the first experiment. The existence of such a rule would not quite solve the free will question completely, as the rule decides the external outcome you experience, not the choice that is made by your brain -- but it would at least point toward a connection between conscious \textit{choices} and physical outcomes that you experience. 



\newpage
\section{{\large SM VI: Statistical Analysis}}
\label{appendix:StatisticalAnalysis}
In this section, we will provide a simple statistical method for identifying if the Born rule is violated. We will see that for some experiments determining Born rule violation is equivalent to the standard problem of determining if a coin is unfair with a low level of false positives and false negatives. 

This can easily be seen for the data for the experimental test of the Redness Rule. In the experiment illustrated in Figure 1, when you repeat the measurement and collect a set of data, you will obtain two numbers: the total number of times you observe red and the total number of times you observe blue. Assuming our experiment is correctly calibrated, then if no Born-rule-violating rule existed, then the probability of seeing blue is $50\%$ and the probability of seeing red is $50\%$. Therefore, the task of our analysis to determine if the Born rule is violated is to determine from the data if these probabilities are not $50\%$ -- which is the same as the classic problem of determining if a coin is fair. 

Therefore well-known methods can be used to determine if data is Born-rule violating. Because of the nature of this experiment requiring you, the reader, to perform the experiment, here we briefly review these well-known techniques to make the potential experiment as accessible as possible. 

 In the next section, we will briefly review this two-tailed binomial p-test, a standard method used to determine if a coin is unfair. Furthermore, we will go over how to calculate the amount of measurements needed to simultaneously have a low false positive and false negative rate for ruling out a certain degree of Born rule violation. This last step is particularly important, because you, the reader, are required to take the data yourself for verifying Born-rule violation. We provide a method that allows testing for multiple different Born-rule violation experiments without the risk of obtaining significant false positives.

\subsection{Review of two-tailed binomial p-test}

Here we briefly review the two-tailed binomial p-test.

Generally speaking, a p-test calculates a p-value, which, given a particular set of data for a particular event, represents the probability of an event ``at least as extreme'' as that event occurring. To see what ``at least as extreme'' means for coin flips, consider the following example: If it is observed that 9 heads occur in 10 flips, then all events at least as extreme as 9 heads are observing 9 and 10 heads. Additionally for a ``two-tailed'' p-test, observing 9 and 10 tails also are included as being ``at least as extreme.'' Therefore the probability of obtaining any of these events is the sum:
\begin{align*}
    \textbf{p} = &Pr_{FB}(H = 9, N = 10) \\
    &+ Pr_{FB}(H = 10, N = 10)\\
    &+ Pr_{FB}(T = 9, N = 10)\\
    &+ Pr_{FB}(T = 10, N = 10),
\end{align*}
\noindent where $Pr_{FB}$ is the probability of obtaining H heads in N flips,
\[
Pr_{FB}(\text{H heads in N flips}) = {N \choose H} \frac{1}{2}^{N}.
\]
This calculated probability is exactly the p-value, and loosely represents the false positive chance, meaning the chance that we would be incorrectly concluding the coin is unfair. 

The general form of the p-value for a fair coin is therefore:
\begin{align*}
\text{p-value} = & \sum_{x=h}^N Pr_{FB}(H = x, N) \\ &+\sum_{x=N-x}^N Pr_{FB}(T = x, N).\\
\end{align*}
In order to know if we should reject the hypothesis that the coin is fair, the typical procedure is to predetermine a ``confidence interval.'' Then if the p-value calculated from experimental data is lower than this confidence interval, the hypothesis that the coin is fair is rejected. 

For reasons we will explain, we recommend first choosing a confidence interval of 5\%. This means that if a calculated p-value is less than $p=.05$, we will conclude that the Born rule is violated. Additionally, having a confidence interval of 5\% means that, in the circumstance that no special Born-rule-violating rules exist, there is a 5\% chance that we obtain a false positive -- and we have inaccurately identified a Born-rule-violating rule when no rule exists. 










We must choose a low enough confidence level to be sure we do not get too many false positives, but the lower we make the confidence level, the more likely that we will obtain false negatives. 

A false negative in our case is the situation when a Born-rule-violating rule exists but the data obtained from it returns a p-value is greater than $p=.05$ -- and consequently we do not conclude that the Born-rule is violated when a Born-rule-violating rule exists. 

For these experiments, having a low false negative rate is significantly more important than having a low false positive rate. This is because we desire to do a single experiment to conclude if or if not the Born rule is violated. If the false negative rate is high, then we cannot know for certain if we have truly eliminated the possibility of a Born-rule-violating rule. 

On the other hand, an initially high false-positive rate is acceptable. Any experiment that appears to violate the Born rule simply needs to then be retested with a lower p-value.

\subsection{Determining False-Positive and False-negative rates}
For a given Born-rule violating rule, significantly more measurements are required to reliably confirm Born-rule violation as the rule more closely resembles the Born-rule's probabilities. For example, confirming the Redness rule for $f = 1$ is significantly easier than $f = .55$, requiring around a thousand times more measurements for the same accuracy. 

Here we describe the technical details to calculate these rates. Specifically, we will show how to determine how much data is required to obtain a sufficiently low false-positive and false-negative rate, given a particular weight of the coin. 

Ideally, assuming we take a sufficiently large set of data, the false-positive rate of the experiment is exactly the confidence interval that is predetermined, which we have chosen as $5\%$. But because of the discrete nature of the Binomial distribution, for smaller sets of data, there is only a discrete number of confidence intervals that can be chosen. This simply means that the experimenter should for a predetermined number of measurements choose a confidence interval that matches one of the possible p-values. 

The false-negative rate depends on the false-positive rate, the fairness of the coin, and on the total number of data points taken. A simulation of the false-negative rate vs these parameters was performed in Python, and is illustrated in Figure 4. As the weight of an unfair coin becomes closer to becoming a fair coin, the more data is required to distinguish it from the fair coin. As the weight of the coin gets closer to a fair 50-50, obtaining a constant false negative rate requires exponentially more coin flips. Consequently, in our experiments, this implies that testing the Redness Rule with probabilities $f = \frac{1}{2}+\epsilon$ and $g = \frac{1}{2}-\epsilon$, as $\epsilon \rightarrow 0$, we require exponentially more data to obtain the same false negative rate. 

\begin{figure}
  \includegraphics[width=\columnwidth ]{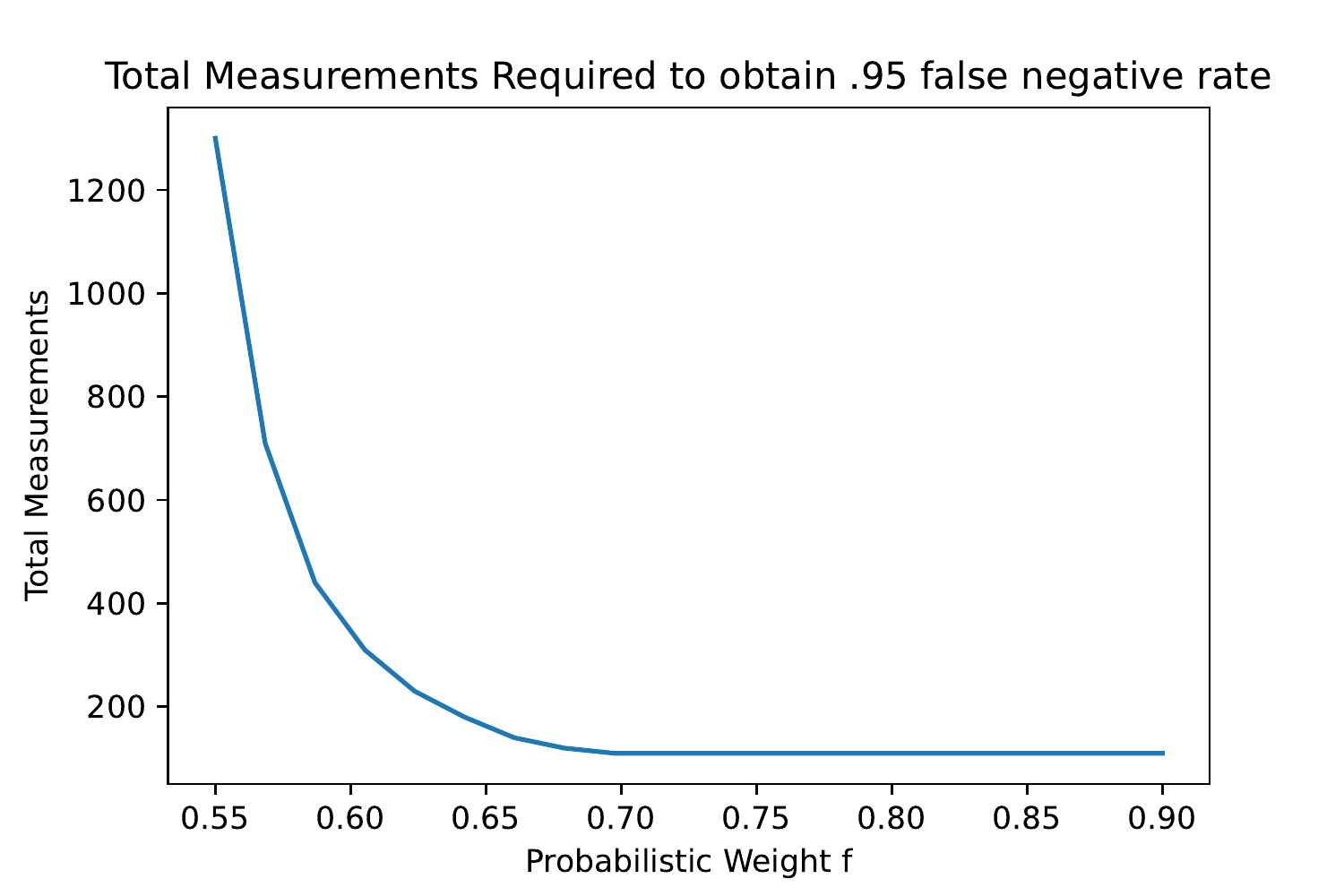}
  \caption{\textbf{Simulation finding the total number of measurements required to obtain a false-negative rate less than 5\%}. In this simulation the false-positive rate is kept near .05, but is adjusted per run for the discrete nature of the binomial distribution. }
   \label{fig:StatsFalseNegative}
\end{figure}

Consequently, a finite amount of measurements can only confirm a subset of rules for the class of the same rule structure with weight f. For the proposed experiments 1, 2 and 4, we recommend taking $\approx 300$ measurements, which gives a false negative rate below 5\% for Born-rule-violating rules that have weights of $f \ge 60\%$, as shown in the Figure \ref{fig:StatsFalseNegative}. For comparison, it would require over 1500 measurements to go from $f \ge 60\%$ to $f \ge 55\%$.


\section{Testing for multiple experiments}
So far we have described the recommended statistical procedure for a single experiment. But as more experiments are added, the chance at at least one of these cases is a false positive becomes more likely. 	For example, there is over a 50$\%$ chance of obtaining at least one false positive after 15 experiments with a 5\% error rate.

In this paper we recommend performing as many experiments as possible to cover as many rules as possible. But performing many experiments will increase number of false positives, which could mislead you into thinking there are Born-rule-violating rules when reality is just statistical fluctuations without the existence any special rules. 

The solution is simple: any values that are measured to have violated the Born rule should be retested to see if it was a false positive. To make this clear that this completely eliminates the increased likelihood of making a false-positive conclusion, We have performed a simple test in Python which conthat false-positives can be virtually be eliminated when any positive results are retested once. We note that this method works largely in part because of the small false-negative rate. The low false negative rate allows us confidently eliminate the possibility of Born-rule violation even after a repeat of an experiment.



\newpage
\section{{\large Appendix I: Wigner's Friend Proof}}

\label{appendix:AppendixWignerFriendProof}

Here we will formally show how Wigner sees his friend in a superposition state of the form:
\[|\psi\rangle =  \tilde{c}_a |F_A\rangle |A\rangle + \tilde{c}_b|F_B\rangle |B\rangle
\]

We begin with a single qubit system (referenced as ``S'') with states $|A\rangle_S$ and $|B\rangle_S$ corresponding to measurements A and B.

The state begins as a superposition of the form: 
\[ \tilde{c}_a |A\rangle_S + \tilde{c}_b |B\rangle_S
\]

When this superposition is measured by Wigner's friend (referenced as ``F'') in the $\{|0\rangle_S, |1\rangle_S \}$ basis, then with probability $|\tilde{c}_a|$ F will measure A with probability $|c_A|^2$ and measure B with probability $|c_B|^2$. And as discussed in the main section, in the friends point of view, the qubit has collapsed to either state $|0\rangle_S$ or state $|1\rangle_S $. 

To Wigner (referenced as ``W''), the closed room that F and S are in is an isolated system. Isolated systems do not have ``wavefunction collapse'' occur and instead evolve unitarily. We can use the laws of unitarity to explicitly find out what this state should be after measurement. The isolated system can be modeled as a product state of the friend and the state: $F\otimes S$. If we instead began with the state $|A\rangle_S$ instead of $\tilde{c}_a |A\rangle_S + \tilde{c}_b |B\rangle_S$, then we know that measuring the state will result in an output product state of the form $|F_A\rangle \otimes |A\rangle_S $ (performing a measurement of state A, returns measurement A). Likewise, if we began with state $|B\rangle_S$ instead, then measuring the state will result in an output product state of the form $|F_B\rangle \otimes |B\rangle_S $. 

Now we know that by the linearity of quantum mechanics, a superposition of these two inputs will produce: 
\[\tilde{c}_a (|F_A\rangle \otimes |A\rangle) + \tilde{c}_b (|F_B\rangle \otimes |B\rangle)
\]

And therefore we see that in Wigner's perspective, the state of the closed room is still an uncollapsed superposition. 

\newpage

\section{{\large Appendix II: Frequently Asked Questions}}
\textbf{Question 1: ``Why should I test for violation of the Born rule when countless experiments have directly and indirectly confirmed that quantum mechanics is correct?''} 

Countless experiments have directly and indirectly confirmed that the Born rule describes our reality, and therefore it is exceedingly essentially impossibly unlikely that the Born rule is violated in a conventional experiment. If someone else, other than you, the reader, observes 1000 quantum coins, it is fundamentally different than if you ,the reader, observe 1000 quantum coins. And if someone else other than you, the reader, performs this experiment, without you, the reader, being there, it is exceedingly, essentially impossibly, unlikely that you will observe that this experimenter observed that the Born rule is violated. Therefore we completely expect that all experiments would observe that the Born rule is upheld.

\textbf{Question 2a: ``But if you and your friend both share the Redness Rule, why is there a disagreement about what happens?''} 

Observer-dependent rules imply there is a difference between which outcome is more likely to be observed in different perspectives. This might appear to cause a contradiction. For example, suppose there is an observer-dependent rule that makes it such that quantum events collapse in a way that make it more likely that you will win the lottery and otherwise follows the Born rule. Therefore in your perspective, you will win the lottery with some high probability. On the other hand, in the perspective of other people, you will not win the lottery. Is that not a contradiction? Either you will have won the lottery or lost the lottery, and after you and any outsider checks the outcome it can only be one of the two outcomes. 

The key misunderstanding here is thinking that there is a single outcome of either winning or losing the lottery. In the many-worlds interpretation, both outcomes occur. When an outside observer checks if you have won the lottery, her wavefunction collapses and they end up in one of these outcomes. Essentially these ``observer-dependent rules" specify which of these universes they end up in.  In this example, you would most likely find yourself in the universe in which you've won the lottery -- but any outside observer, in her perspective, would most likely find themselves in a universe in which you have not won the lottery. This difference is because of the rule we have specified, that collapse events will be biased in favor of the observer winning the lottery. Unless that outside observer collapses a quantum event that causes \textit{them} to win the lottery, quantum events collapse normally for them. 

 
 \textbf{Question 2b: ``In the `Redness Rule' example, observers are more likely to see red events. So, if my friend does this experiment, he should never see blue. But in my perspective as an outside observer, the rule does not apply to me and my observation of him follows the Born rule -- and therefore I can observe my friend observing blue. Does not that contradict this supposed `Redness Rule' that he can only see red?''} 
 
 This is the same misunderstanding as Question 2a. 
 
 In your perspective, either option happens, but in your friend's perspective the option ``red'' happens more often. There is no contradiction because it just describes which of the many worlds each observer independently finds themselves in. There is a world where your friend see's red and a world where your friend see's blue. He always finds himself in the universe where he sees red, and you find yourself in either universe because the condition of the rule does not apply to you, since you collapse the state by observing your friend, \textit{not} by observing an outcome of red or blue. 
 

 \textbf{Question 3: ``What is the point of collecting a bunch of people and seeing if they are more likely to see certain colors? I have no reason to believe this test would find anything.''} 
 
This is very much not what is being proposed. It is true that the test for the observer-dependent ``Redness Rule'' requires observers to collapse a quantum event by observing one of two colors. But the key idea is that even if this ``Redness Rule'' existed, performing a standard experiment would not find anything unexpected. If a thousand people are collected to perform a measurement in the Redness Rule experiment, even if the ``Redness Rule'' is correct, $\approx 500$ will observe red and $\approx 500$ will observe blue -- and the results will follow the expected Born rule.  

All observer-dependent rules will follow the Born rule, unless you, the reader, are the person performing the experiment. Anyone else who performs this experiment will simply get results expected by the Born rule.

\textbf{Question 4: There already is an infinite set of untested laws of physics. Many of these untested rules are completely arbitrary. For example, have we tested for the possibility that people sneezing causes the strength of the electromagnetic force across the entire universe to change? Your proposed set of theories seem just as arbitrary. } 

In this arbitrary example of sneezing changing the strength of the electromagnetic force, the scientific community could do an experiment and prove that it does not exist. The point is that there's nothing about this law that prevents the scientific community from discovering it. So it is reasonable to believe that lack of discovery of anything of this sort is evidence that it does not exist. Humanity has existed for a very long time, and arbitrary laws of physics have had a long time to have been discovered.

Essentially, anything like this arbitrary rule is just challenging the scientific system in a completely conventional way. This paper is not meant to be a philosophy of science paper, and we will assume that these types of questions have already been sufficiently answered as standard epistemology. We agree that these sort of arbitrary rules are not interesting to consider.

But what is unique about this proposal is that \textit{these observer-dependent rules cannot be deduced by outside observers, other than you the reader.} Therefore, it is impossible for outside observers to have observed these laws. Essentially, unlike most arbitrary laws of physics which have the \textit{potential} to be ruled out, these observer-dependent rules are undiscoverable by conventional scientific experiments. 

These experiments can only be confirmed if you, the reader, performs the experiment. This is why it is distinct from just adding some arbitrary new law of physics - because arbitrary new law of physics could have been tested by someone else.

\textbf{Question 5: Well then this is just some trick. This has nothing to do with quantum mechanics and everything to do with just stipulating that there are ``observer-dependent'' outcomes. } 

It is not obvious how there is a classical equivalent of such ``observer-dependent'' rules without blatant contradictions. This quantum formalism avoids inconsistencies because of the many worlds. Outcomes can depend on the observer because it simply describes which universe the observer is in. In a classical formulation of an ``observer dependent rule,'' there is only one universe, so if any new laws of physics are added that dependent on the point of view of the observer, that rule would have to resolve how two observers can be in disagreement. Relativity is an example of a classical formulation of an observer dependent rule, as observers can have contradictory measurements of the same thing in their point of view. It is possible that there are more ``observer dependent rules'' in a classical setting, but it is not obvious to us if it is possible to formulate these rules in a way that is not contradictory. 

For example, consider the rule that ``the strength of the electric force changes in my perspective when people sneeze.'' Here we have modified our arbitrary rule from question 4 to be an observer-dependent rule. Now we're stipulating that the force changes \textit{in my perspective}. If it does one thing in one perspective and something else in another person's perspective, how is this contradiction resolved? In this case I would observe a force changing and others would not. This is very different from the quantum case as it requires that observers will disagree on the most basic things. Similar to relativity, such a theory would need to resolve why there is a blatant disagreement between observers. But it is in fact a very interesting question that could potentially be explored: other than relativity, are there other "observer-dependent" rules that can consistently exist? 

\textbf{Question 6: If it cannot be tested then it is not science.} 
It certainly can be tested. This paper proposed a set of real experiments you can do to test these observer-dependent laws. 

\textbf{Question 7: If it cannot be tested \textit{by the scientific community} then it is not science.} 
Each individual \textit{in the scientific community} can perform this test on his or her own. 

Additionally, we mention that if \textit{you} the reader choose not to perform a test for this, can you really justify under the premise that the test is not scientific -- when in performing the experiment you are finding out evidence of the nature of your reality? 

This experiment might demonstrate that such a definition of what is ``science'' is not great for \textit{you} to discover the truth about \textit{your} reality, which is perhaps another way to think of what ``science'' is.

\end{appendices}


{}

\end{document}